\documentclass[iop]{emulateapj}
\usepackage{amsmath}
\usepackage[caption=false]{subfig}
\usepackage[colorlinks=true,citecolor=blue]{hyperref}
\newcommand{\phat}[1]{\expandafter\hat#1}
\begin{document}

\title{Delay Spectrum with Phase-Tracking Arrays: Extracting the
HI power spectrum from the Epoch of Reionization}
\author{Sourabh Paul\altaffilmark{1}, Shiv K. Sethi\altaffilmark{1}, Miguel F. Morales\altaffilmark{2}, K. S. Dwarkanath\altaffilmark{1}, N. Udaya Shankar\altaffilmark{1},  Ravi Subrahmanyan\altaffilmark{1},  N. Barry\altaffilmark{2}, A. P. Beardsley\altaffilmark{3}, Judd D. Bowman\altaffilmark{3}, F. Briggs\altaffilmark{4,5}, P. Carroll\altaffilmark{2}, A. de Oliveira-Costa\altaffilmark{6}, Joshua S. Dillon\altaffilmark{12}, A. Ewall-Wice\altaffilmark{6}, L. Feng\altaffilmark{6}, L. J. Greenhill\altaffilmark{7}, B. M. Gaensler\altaffilmark{5,8,17}, B. J. Hazelton\altaffilmark{2}, J. N. Hewitt\altaffilmark{6}, N. Hurley-Walker\altaffilmark{9}, D. J. Jacobs\altaffilmark{3}, Han-Seek Kim\altaffilmark{5,10}, P. Kittiwisit\altaffilmark{3}, E. Lenc\altaffilmark{5,8}, J. Line\altaffilmark{5,10}, A. Loeb\altaffilmark{7}, B. McKinley\altaffilmark{5,10}, D. A. Mitchell\altaffilmark{5,11}, A. R. Neben\altaffilmark{6}, A. R. Offringa\altaffilmark{4,5}, B. Pindor\altaffilmark{5,10}, J. C. Pober\altaffilmark{19}, P. Procopio\altaffilmark{5,10}, J. Riding\altaffilmark{5,10}, I. S. Sullivan\altaffilmark{2}, M. Tegmark\altaffilmark{6}, Nithyanandan Thyagarajan\altaffilmark{3}, S. J. Tingay\altaffilmark{5,9,18}, C. M. Trott\altaffilmark{5,9}, R. B. Wayth\altaffilmark{5,9}, R. L. Webster\altaffilmark{5,10}, J. S. B. Wyithe\altaffilmark{5,10}, Roger Cappallo\altaffilmark{13}, M. Johnston-Hollitt\altaffilmark{14}, D. L. Kaplan\altaffilmark{15}, C. J. Lonsdale\altaffilmark{13}, S. R. McWhirter\altaffilmark{13}, E. Morgan\altaffilmark{6}, D. Oberoi\altaffilmark{16}, S. M. Ord\altaffilmark{5,9}, T. Prabu\altaffilmark{1}, K. S. Srivani\altaffilmark{1}, A. Williams\altaffilmark{9}, C. L. Williams\altaffilmark{6}}

\altaffiltext{1}{Raman Research Institute, Bangalore 560080, India}
\altaffiltext{2}{Department of Physics, University of Washington, Seattle, WA 98195, USA}
\altaffiltext{3}{School of Earth and Space Exploration, Arizona State University, Tempe, AZ 85287, USA}
\altaffiltext{4}{Research School of Astronomy and Astrophysics, Australian National University, Canberra, ACT 2611, Australia}
\altaffiltext{5}{ARC Centre of Excellence for All-sky Astrophysics (CAASTRO)}
\altaffiltext{6}{Kavli Institute for Astrophysics and Space Research, Massachusetts Institute of Technology, Cambridge, MA 02139, USA}
\altaffiltext{7}{Harvard-Smithsonian Center for Astrophysics, Cambridge, MA 02138, USA}
\altaffiltext{8}{Sydney Institute for Astronomy, School of Physics, The University of Sydney, NSW 2006, Australia}
\altaffiltext{9}{International Centre for Radio Astronomy Research, Curtin University, Bentley, WA 6102, Australia}
\altaffiltext{10}{School of Physics, The University of Melbourne, Parkville, VIC 3010, Australia}
\altaffiltext{11}{CSIRO Astronomy and Space Science (CASS), PO Box 76, Epping, NSW 1710, Australia}
\altaffiltext{12}{Departments of Physics and Astronomy, University of California, Berkeley, CA, 94720, USA}
\altaffiltext{13}{MIT Haystack Observatory, Westford, MA 01886, USA}
\altaffiltext{14}{School of Chemical \& Physical Sciences, Victoria University of Wellington, Wellington 6140, New Zealand}
\altaffiltext{15}{Department of Physics, University of Wisconsin--Milwaukee, Milwaukee, WI 53201, USA}
\altaffiltext{16}{National Centre for Radio Astrophysics, Tata Institute for Fundamental Research, Pune 411007, India}
\altaffiltext{17}{Dunlap Institute for Astronomy and Astrophysics, University of Toronto, ON M5S 3H4, Canada}
\altaffiltext{18}{Istituto Nazionale di Astrofisica (INAF) \textemdash Istituto di Radio Astronomia, Via Piero Gobetti, Bologna, 40129, Italy}
\altaffiltext{19}{Department of Physics, Brown University, Providence, RI 02912, USA}

\begin{abstract}
The Detection of redshifted  21 cm emission  from the epoch of 
reionization  (EoR) is a challenging task owing to strong foregrounds that dominate the signal.  In this paper, we propose a general 
 method, based on the delay spectrum approach,  to extract HI power spectra that is applicable to
 tracking observations using
an imaging  radio interferometer (Delay Spectrum with Imaging Arrays (DSIA)). 
Our method is based on modelling the HI signal taking
into account the impact of wide field effects such as the  $w$-term  which
  are then used  as  appropriate weights in  cross-correlating  the measured visibilities. Our method is applicable to any   
radio interferometer that tracks a phase center and could be 
utilized for arrays such as   MWA, LOFAR,  GMRT, PAPER and HERA.  In the literature the delay spectrum approach has been  implemented for near-redundant baselines using  drift scan observations. 
In this paper we explore the scheme for non-redundant tracking arrays, and this is the first application of delay spectrum methodology  to such
data to extract the HI signal.
We analyze 3 hours of MWA tracking data on the EoR1 field. We present both 2-dimensional ($k_\parallel,k_\perp$) and 1-dimensional (k) power spectra from the analysis. Our results 
are in agreement with the findings  of other pipelines 
developed to analyse the MWA EoR data. 

\end{abstract}
\maketitle
\section{Introduction}
The probe of the Epoch of Reionization (EoR) remains one of the outstanding 
aims of modern cosmology. In the past decade, intriguing details have emerged
about this epoch from a host of cosmological observables. Gunn-Peterson (GP)
tests on spectra of quasi-stellar objects (QSOs) ((\citet{Fan et al}) in the redshift range $5.7< z <6.3$ suggest that the universe 
as making a transition to full reionization during this period. On the 
other hand, Cosmic microwave background radiation (CMBR) temperature and polarization anisotropy measurements (\citet{Komatsu et al}, \citet{Planck 2015}) suggest that the universe might have
been fully ionized in a redshift range $8 < z < 10$ (\citet{Planck 2016}). Both these observables
have their strengths and weaknesses. The GP test, based on Lyman-$\alpha$ 
absorption, is not able to distinguish between a fully neutral medium from
the one ionized to one part  in a thousand. CMBR anisotropies, based on 
photon scattering off free electrons, are  sensitive to the integrated 
Thompson scattering optical depth and therefore cannot reliably 
construct the  tomography of the reionization epoch. 

In the recent past, major experimental efforts have been undertaken to 
study the EoR  in  redshifted 21-cm line emission from the epoch. In particular, 
many radio interferometers in frequency range $80 < \nu < 300 \, \rm MHz$ 
are currently operational that specifically  aim to detect the EoR, for example, Low Frequency Array (LOFAR, \citet{LOFAR}), 21 Centimeter Array (21CMA, \citet{21cma}), Giant Meterwave Radio Telescope (GMRT, \citet{GMRT}), Donald C. Backer Precision Array for Probing the Epoch of Reionization (PAPER, \citet{PAPER}), and the Murchison Widefield Array (MWA, \citet{MWA, Lonsdale et al, Bowman et al 2009}). Even though 
the detection of redshifted HI line from the EoR remains the most direct
and,  possibly the most promising, way to delineate the details of the epoch, 
this method is beset with its own set of issues. First, unlike 
 CMBR anisotropies, the theoretical modelling of the HI signal from the 
 EoR is considerably harder principally  owing to uncertainty in the nature of ionizing sources and the details of their formation and evolution. Second, 
the signal is expected to be weak with brightness temperature $T_{\rm B} \simeq 10 \, \rm mK$. Many hundred hours of observation is needed to detect such a 
signal with  current interferometers, with the attendant  complication of 
maintaining  instrumental stability for  such long periods.  Third, the foreground are expected to be many orders of magnitude larger than the signal (for details on the three issues see e.g.  \citet{Morales Wyithe, Morales Hewitt, Furlanetto et al, Barkana 2001, Zaroubi 2013} and references therein). 

The use of  radio interferometers 
 to  estimate   the underlying power spectra has been successfully
employed for CMB data analysis (\citet{Hobson CMB}). This method has also
been suggested as a possible probe of the intensity correlations  of the 
redshifted HI line, including from the EoR (\citet{Bharadwaj and Sethi, Datta et al, Bharadwaj and Ali}). 

Many different approaches have been discussed to detect the HI signal in the 
presence of dominant foregrounds \citet{Hazelton foreground, Jelic 2008, Harker 2009, Liu 2011, Morales 2012, Trott 2012, Dillon 2013, Delay 3}. They are all based on the expectation
that  foregrounds are  smooth in frequency space  as they arise from continuum
emission, e.g. Synchrotron radiation,  in both our Galaxy and extra-galactic 
sources. On the other hand the HI signal has significant structure in 
 the frequency space. It is conceivable that all these sources, both point and diffuse,
can be subtracted from the images, leaving behind the HI signal and Gaussian 
noise, and LOFAR partly relies upon this technique (\citet{Chapman 2012, Chapman 2013}). 
 Another possible method is based on the isolation of foregrounds
from the HI signal using power spectrum of the observed signal in conjugate space to the observed frequency (\citet{Parsons 2012a, Delay 1, Pober PAPER, Thyagrajan et al}).  Variants of this `delay space' (\citet{Parsons 2012a, Delay 1}) method  are  particularly relevant for interferometers such as MWA that have low angular resolution and have been 
used extensively for the analysis of PAPER data. In this approach the observed interferometric data---visibilities for each antenna pair as a function of 
frequency---is Fourier transformed along the frequency axis. The Fourier conjugate variable  effectively captures signal delay between  antenna pairs, which  allows one to isolate foregrounds.  In  the context 3d HI power spectrum this variable can  be related to  cosmological distance along line of sight (for details of this  mathematical correspondence see  \citet{Delay 1, Delay 2}). The `delay spectrum' constructed from `delay space' approach can be used to recover the  cosmological 3d  HI power spectrum. In this approach,    one deals  with visibilities directly, which are primary data products of  radio interferometers.

In this paper we propose a new method based on the delay space approach  to 
extract the power spectrum of the HI signal in the presence of noise and 
foregrounds. Our method  is based on  modelling the HI signal taking 
into account the impact of  the $w$-term (arising from  non-coplanarity of the array, \citet{W projection}) and the distortion of  intensity pattern
during a tracking run. The information from  the HI signal
is used as weights  to cross-correlate the measured visibilities.  The proposed 
method (`Delay Spectrum with Imaging Arrays (DSIA)')  
 is a general method  applicable for tracking  with 
  radio interferometers with wide  primary beams and  arbitrary array configuration  (e.g. MWA, LOFAR) and can also be 
applied to interferometers with redundant baselines.  We apply the proposed 
 method to analyse 3~hours of  MWA data  and compare our
results with noise and foreground simulations.

The outline of this paper is as follows. In the next 
section, we describe the method of visibility correlation using 
delay space  approach in detail. In particular, 
this method is applied to the HI signal. In section~3, we describe  the
 MWA data and the initial analysis of this data based on the publicly-available image processing 
software: Common Astronomical Software Applications (CASA). In section~4   the pipeline to extract the HI 
power spectrum is discussed. In section~5, the results from MWA data are 
 discussed and compared with  simulations of foregrounds
and noise. In the final section, we summarize our results and indicate possible
future directions. 

Throughout this paper, we have used the Planck+WP best fit values of
cosmological parameters: $\Omega_{\rm m0}=0.3183$,
$\Omega_{\rm \Lambda0}=0.6817$ , $\Omega_{\rm b0}\,h^2=0.02203$, 
$h=0.6704$, $\sigma_8=0.8347$, and $n_{\rm s}=0.9619$ 
\citet{Planck 2015, Planck 2013}.

\section{HI signal and its correlations}

\label{sec:h1signal}

In this section we study the HI signal using visibility correlations in
delay space. Using 
our formulation we derive, in addition to new results, 
 many results already known in the literature. The main new results
are: the impact of $w$-term  (subsection~2.1) and 
changing intensity pattern in a tracking run  (subsection~2.2) on the 
HI correlations.  Our results are valid for any radio interferometer
but our aim here is to underline their applicability  to MWA. 

The most important inputs from MWA array configuration for our study are: 
(a) the MWA primary beam and (b) the bandwidth. Other properties of the MWA array that have a bearing on our analysis are the non-coplanarity of the array
 and its
baseline distribution.

In delay space, spectrally smooth foregrounds lend themselves to ready 
interpretation.  It can be shown that  visibilities computed in delay 
space allow isolation of such foregrounds from the regions dominated 
by the EoR signal and noise  (e.g. see \citet{wedge 1, wedge 2, Parsons 2009, Parsons 2012a, Delay 1, Delay 2, Liu 2014b, Delay 3, Thyagrajan et al, Delay 4}).
This can be achieved by Fourier transforming the raw visibilities in frequency space.  
However, being 3-dimensional and  statistical in nature, the  properties of the HI signal can 
only be inferred  by correlating the observed visibilities. Our approach, which is 
based on visibility correlations in delay space,   allows us to develop a unified method  to deal with both the HI 
signal and the foregrounds, which  are discussed in Appendix~\ref{sec:foregrounds}.

Radio interferometers measure the spatial correlation of the electric fields from the sky, the visibility $V_\nu({\bf u}_\nu)$:
\begin{equation}
 V_\nu({\bf u}_\nu)=\int A_\nu(\vec{\theta}) I_\nu (\vec{\theta})e^{-i2\pi {\bf u}_\nu \cdot{\vec{\theta}}} d^2\theta \label{visibility}
\end{equation}
Here ${\bf u}_\nu=(u_\nu,v_\nu)$ is the distance vector between the antennas of the interferometer (also called baseline vector) in units of wavelength, $\vec{\theta}$ denotes the position vector on the sky plane (which can be be expressed 
as a pair of direction cosines $\{l,m\}$)  and $\nu$ is the frequency of observation. 

We neglect the impact of $w$-term in this section in writing the relation
between the visibility and specific intensity in Eq.~(\ref{visibility}). The $w$-term arises from 
non-coplanarity of the interferometric array and changes as the   interferometer
tracks a region.
In section~\ref{sec:h1sigwterm} we show  that the inclusion of $w$-term causes an effective shrinking of the primary beam.   For our study  we  calculate how  the HI signal is affected owing  to  non-zero $w$  as compared to $w=0$ case.

We explicitly express the frequency dependence of all the quantities. These quantities are: the background
specific intensity $\bar{I_\nu}$, the primary beam $A_\nu(\vec\theta)$ and the 
baseline ${\bf u}_\nu$.  In terms of brightness temperature $T_B$, $I_\nu (\vec{\theta}) = 2kT_{B}(\vec{\theta})/\lambda^2$. 

 The sky intensity can be decomposed as: 
\begin{equation}
 I_\nu(\vec\theta)=\bar{I}_\nu + \Delta I_\nu(\vec\theta)
\end{equation}
where $\bar{I}_\nu$ and $\Delta I_\nu(\vec\theta)$ are the isotropic and fluctuating parts of the intensity distribution. Since the isotropic component does not contribute to interferometric measurement \footnote{However some methods have been discussed in recent literatures to extract monopole signal from interferometric measurements (\citet{Presley 2015, Singh 2015}).}, the visibility recorded at frequency $\nu$ can be written as:
\begin{equation}
 V_\nu({\bf u}_\nu)=\int A_\nu (\vec\theta) \Delta I_\nu(\vec\theta)e^{-i2\pi\vec\theta.{\bf u}_\nu} d^2\theta
 \label{vis_a}
\end{equation}

The measured visibility receives contributions from the redshifted HI line,
the foregrounds, and the noise. 

For the HI signal the observed intensity fluctuations can be 
related to the HI perturbations in Fourier space, $\delta_{\rm HI}({\bf k})$, as:
\begin{equation}
 \Delta I_\nu(\vec\theta)= \bar{I}_\nu\int\frac{d^3k}{(2\pi)^3}\delta_{\rm HI}({\bf k})
 e^{i{\bf k}\cdot {\bf r}}
 \label{deli_a}
\end{equation}
Here ${\bf r} =\{\vec\theta,r_\nu\}$ specifies the three-dimensional position of 
the the HI emission; $r_\nu$ is  the coordinate distance to the point of observation: $r_\nu = \int dz/H(z)$ with the limit of this integral extending from zero to redshift  $z = \nu_e/\nu_0 - 1$. 
$\delta_{\rm HI}({\bf k})$  comprises of many physical effects: density fluctuations, ionization inhomogeneity, density-ionization fraction cross-correlation, etc. (\citet{Furlanetto et al, Zaldarriaga 2004}). Together this can be expressed as:
\begin{equation}
 \delta_{\rm HI}({\bf k})=\beta_b\delta_b+\beta_x\delta_x+\beta_\alpha\delta_\alpha+\beta_T\delta_T-\delta_{\partial v}
\end{equation} 
Each quantity in the above equation corresponds to the fractional variation of a particular physical quantity:  $\delta_b$ refers to fluctuation in baryonic density,
$\delta_\alpha$ for the Ly$\alpha$ coupling coefficient $x_\alpha$, $\delta_x$ for the neutral fraction, $\delta_T$ for $T_K$, and 
$\delta_{\partial v}$ for the line of sight peculiar velocity gradient. $\beta$ factors denote the expansion coefficients of the corresponding quantity (\citet{Furlanetto et al}).

Current experiments such as MWA, LOFAR and PAPER aim 
statistical detection of the EoR signal. The quantities of interest here 
are the correlation functions of the HI fluctuations. The most important 
correlation function one seeks to detect in an EoR experiment
is  the power spectrum, $P_{\rm HI}$, defined as:
\begin{equation}
 \langle\delta_{\rm HI}^*({\bf k})\delta_{\rm HI}({\bf k'})\rangle=(2\pi)^3\delta^3({\bf k}-{\bf k'})P_{\rm HI}(k)
\label{powspec3d}
\end{equation}
 The HI power spectrum can be constructed from the correlation of the 
observed visibilities.  Substituting the form of fluctuation $\Delta I_\nu$ from Eq.~(\ref{deli_a}) in the visibility expression (Eq.~(\ref{vis_a})), we get:
\begin{eqnarray}
 V_\nu({\bf u}_\nu)&=&\bar{I}_\nu\int\frac{d^3k}{(2\pi)^3}\delta_{\rm HI}({\vec {k}})e^{ir_\nu k_\parallel}
 \int  A_\nu(\vec\theta) \nonumber \\
 &&\times\exp\left[-2\pi i\left({\bf u}_\nu-\frac{{\bf k_\perp}r_\nu}{2\pi}\right)\ldotp\vec{\theta}\right]d^2\theta 
\end{eqnarray}
Here we have decomposed the wave vector ${\bf k}$ as  
components on the plane of the sky, ${\bf k}_\perp$ and along the line of 
sight, $k_\parallel$. The integral over angles  is  the Fourier transform of the primary beam $A_\nu(\vec{\theta})$, which allows us to re-write this equation as:
\begin{equation}
 V_\nu({\bf u}_\nu)=\bar{I_\nu}\int\frac{d^3k}{(2\pi)^3}\delta_{\rm HI}({\bf k})
 e^{ir_\nu k_\parallel}a\left({\bf u}_\nu-\frac{{\bf k_\perp}r_\nu}{2\pi}\right)
\label{visdef}
\end{equation}
where 
\begin{eqnarray}
a\left({\bf u}_\nu- \frac{{\bf k_\perp}r_\nu}{2\pi}\right)&& \equiv \int  A_\nu(\vec\theta) \nonumber \\
&&\times\exp\left[-2\pi i\left({\bf u}_\nu-\frac{{\bf k_\perp}r_\nu}{2\pi}\right)\ldotp\vec{\theta}\right]d^2\theta
\end{eqnarray}

Using Eq.~(\ref{powspec3d}), the visibility correlation function can 
be computed:
\begin{eqnarray}
\langle V_\nu({\bf u}_\nu) V_{\nu'}^* &&({\bf u}'_{\nu'}) \rangle = \bar{I_\nu}^2\int\frac{d^3k}{(2\pi)^3}P_{\rm HI}(k) e^{i\Delta r_\nu k_\parallel}\nonumber \\
&&\times a\left({\bf u}_\nu-\frac{{\bf k_\perp}r_\nu}{2\pi}\right) a\left({\bf u'}_\nu-\frac{{\bf k_\perp}r_{\nu'}}{2\pi}\right) 
\label{viscorrfreq}
\end{eqnarray}
Here $\Delta r_\nu = |r_{\nu'}-r_\nu|$ which for $|\nu'-\nu| \ll \nu$ can be simplified to: $\Delta r_\nu =r_\nu'|\nu'-\nu|$, where $r_\nu' = |dr_\nu/d\nu|$. 
Eq.~(\ref{viscorrfreq}) gives the correlation of the HI signal in three dimensions in which the two coordinates ${\bf u}$ correspond to  Fourier components
of the HI signal while the third $\nu$ refers to  the coordinate of 
the fluctuation in position,  $r_\nu$, space  (\citet{Bharadwaj and Sethi}).

To isolate the impact of foregrounds and obtain regions dominated by 
the HI signal  and the noise (`EoR window'), we compute the 
the visibilities in delay space (\citet{Parsons 2009, Parsons 2012a, Delay 1, PAPER, Delay 2}):
\begin{equation}
V_\tau({\bf u}) =  \int \exp(i2\pi \tau \nu) V_\nu({\bf u}_\nu) d\nu
\label{deftau}
\end{equation}
Here $\tau$, the conjugate variable of $\nu$, defines the relevant variable
in delay space. The delay space approach can be applied to 
 data to isolate spectrally smooth foregrounds; we discuss the delay space approach as applied to
such foregrounds in Appendix~A.  In Eq.~(\ref{deftau}), we have suppressed the frequency dependence of the baseline on
the LHS as  the frequency dependence of all the quantities has been integrated.
The baseline vector can be expressed as: ${\bf u}_\nu = {\bf u}_{\nu_0} \nu/\nu_0$,where $\nu_0$ is some fixed frequency that lies within  the bandwidth. On the 
LHS of Eq.~(\ref{deftau}), the frequency independent baselines,  ${\bf u} \equiv {\bf u}_{\nu_0}$. Throughout this paper, we  assume: $\nu_0 = 154 \, \rm MHz$,  the central frequency of the bandwidth we use for MWA data analysis.

The autocorrelation of $V_\tau({\bf u})$ can be written as: 
\begin{eqnarray}
&&\langle V_\tau({\bf u}) V_\tau ^* ({\bf u'}) \rangle  =  \int  \int d\nu d\nu'  \bar{I_\nu}\bar{I_{\nu'}} \int {d^3k \over (2\pi)^3} P_{\rm HI}(k)\nonumber \\
&\times & \exp\left(i2\pi\tau(\nu'-\nu)+i(k_\parallel(r_{\nu'}-r_\nu) \right) \nonumber \\
& \times &  \int d^2\theta  \exp\left(i2\pi\vec\theta\cdot\left({\bf u}_\nu-\frac{{\bf k_\perp}r_\nu}{2\pi}\right) \right) A_\nu(\vec\theta)\nonumber \\
&\times &\int d^2\theta' \exp\left(- i2\pi\vec\theta'\cdot\left({\bf u'}_{\nu'}-\frac{{\bf k_\perp}r_{\nu'}}{2\pi}\right)\right) A_{\nu'}(\vec\theta') 
\label{etacorrela}
\end{eqnarray}
 To make further progress, frequency 
dependent quantities are Taylor expanded. For baselines, this is 
a straightforward  re-expression of the baseline vector as the vector is 
linear in frequency:   ${\bf u'}_{\nu'} = {\bf u'}_\nu +\Delta\nu d{\bf u'}_\nu/d\nu$ where $\Delta\nu = \nu'-\nu$. It should be  noted that $d{\bf u'}_\nu/d\nu$ is the physical baseline length measured in the  units of time.

After the Taylor expansion of relevant quantities, $r_\nu$ and 
${\bf u}_\nu$, we obtain:
\begin{eqnarray}
&&\langle V_\tau({\bf u}) V_\tau ^* ({\bf u'}) \rangle  =  \int  \int d\nu d\nu'  \bar{I_\nu}\bar{I_{\nu'}}   \int {d^3k \over (2\pi)^3} P_{\rm HI}(k)\nonumber \\
&\times & \int d^2\theta  \exp\left(i2\pi\vec\theta\cdot\left({\bf u}_\nu-\frac{{\bf k_\perp}r_\nu}{2\pi}\right) \right) A_\nu(\vec\theta) \nonumber \\
& \times & \int d^2\theta' \exp\left(- i2\pi\vec\theta'\cdot\left({\bf u'}_{\nu}-\frac{{\bf k_\perp}r_{\nu}}{2\pi}\right)\right) A_{\nu'}(\vec\theta')\nonumber \\
&\times & \exp\left[i\Delta\nu \left(2\pi\tau+k_\parallel dr_\nu/d\nu + d{\bf u'}_\nu/d\nu\cdot\vec\theta'\right. \right. \nonumber \\
&+ &  \left.\left.   dr_\nu/d\nu  {\bf k}_\perp\cdot\vec\theta'/(2\pi)  \right)\right]
\label{etacorrela12}
\end{eqnarray}
Here $\Delta\nu = \nu'-\nu$;  all the quantities in Eq.~(\ref{etacorrela12}) 
 have been written
as explicit functions of $\nu$ and $\Delta\nu$. This allows us to simplify the 
integral further by making the coordinate transform $y = (\nu+\nu')/2$ and
$x= (\nu'-\nu)/2$; the Jacobian of this transformation is unity. We can make
further simplification by using  $y \simeq \nu$. This is justified for our case 
as we assume the bandwidth  to be $\simeq 10 \, \rm MHz$ around a central
frequency of $\simeq 150 MHz$. All the frequency dependent variables change by
less than $10\%$ for this case. This 
reduces  Eq.~(\ref{etacorrela12}) to: 
\begin{eqnarray}
&&\langle V_\tau({\bf u}) V_\tau ^* ({\bf u'}) \rangle  =  \int  \int d\nu dx  \bar{I_\nu}^2   \int {d^3k \over (2\pi)^3} P_{\rm HI}(k) \nonumber \\
&\times &\int d^2\theta  \exp\left(i2\pi\vec\theta\cdot\left({\bf u}_\nu-\frac{{\bf k_\perp}r_\nu}{2\pi}\right) \right) A_\nu(\vec\theta) \nonumber \\
& \times & \int d^2\theta' \exp\left(- i2\pi\vec\theta'\cdot\left({\bf u'}_{\nu}-\frac{{\bf k_\perp}r_{\nu}}{2\pi}\right)\right) A_\nu(\vec\theta')\nonumber \\ &\times & \exp\left[i2x \left(2\pi\tau+k_\parallel dr_\nu/d\nu + d{\bf u'}_\nu/d\nu\cdot\vec\theta'\right. \right. \nonumber \\
&+ &  \left.\left.  dr_\nu/d\nu  {\bf k}_\perp\cdot\vec\theta'/(2\pi)  \right)\right]
\label{etacorrela1}
\end{eqnarray}
Given the HI power spectrum $P_{\rm HI}(k)$ this integral could be 
computed numerically. However, it is possible to determine the correlation scales
in both the transverse and line of sight directions by carefully examining  
Eq.~(\ref{etacorrela1}). The integral over angles  shows that the dominant
contribution comes from wavenumbers such that ${\bf k}_\perp \simeq  2\pi {\bf u}_\nu/r_\nu$. This relation allows us to simplify the integrals over $\theta$, $\theta'$ and $x$. In particular, different terms in the 
 exponent containing $\tau$ can be estimated. Using  $d{\bf u}_\nu/d\nu = {\bf u}_\nu/\nu$,  the last two terms
in the exponents are  on the order of ${\bf k}_\perp.\vec\theta' r_\nu/(2\pi\nu)$ (the term containing $dr_\nu/d\nu$ is slightly smaller because $\nu/r_\nu dr_\nu/d\nu  \simeq 0.3$). For MWA primary beam,  $\theta_0 \simeq 0.3$, and 
for MWA baseline distribution, the term $k_\parallel dr_\nu/\nu$ generally dominates over these terms, especially in the regions dominated by EoR. For all our calculations we use parameters specific to MWA, in particular, the primary beam of MWA. However, the formulation presented here is general enough 
to be applicable to other arrays.

\begin{figure}
\centering
\includegraphics[width=0.5\textwidth]{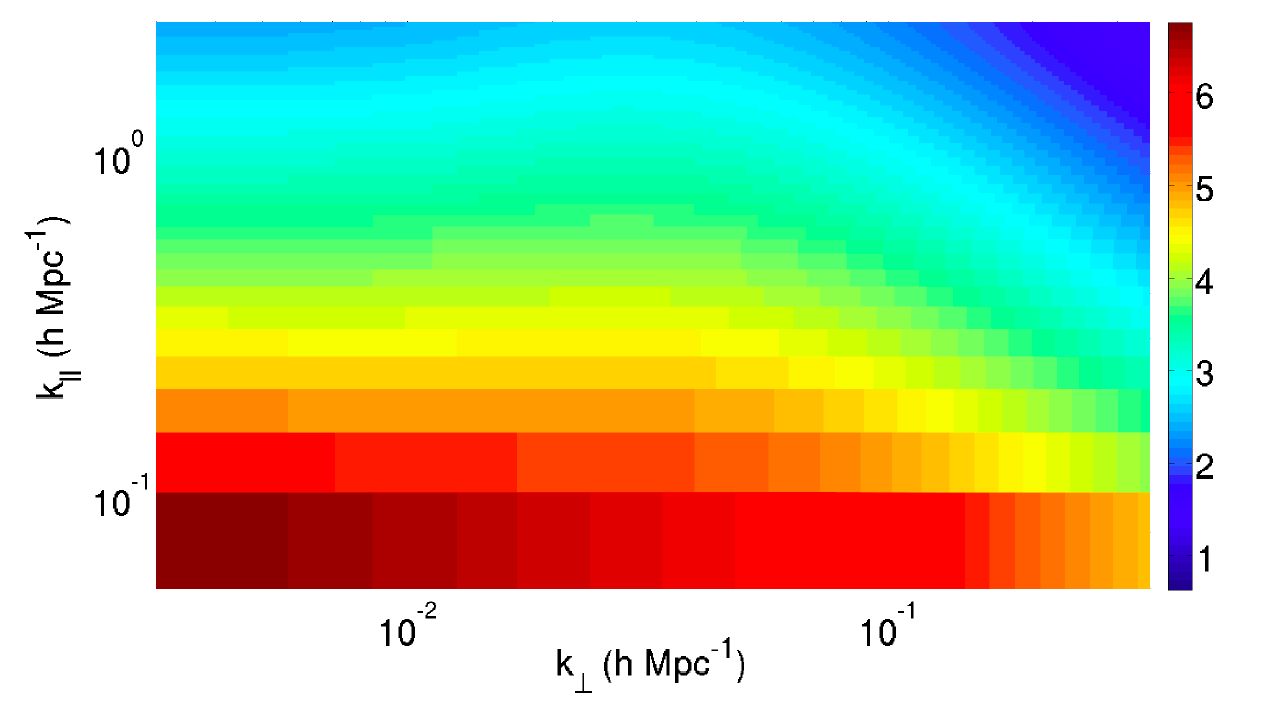}
\caption{Power spectra of the expected Cosmological HI signal. The power is plotted as $log_{10}P$ where P is in units  $mK^2 (Mpc/h)^3$.}
\label{signal_ps}
\end{figure}

By dropping the last  two  terms, which are subdominant,  in the exponent containing $\tau$, we can 
separate the integrals over $x$ and angles, this gives us:
\begin{eqnarray}
&&\langle V_\tau({\bf u}) V_\tau ^* ({\bf u'}) \rangle  =  \int   d\nu   \bar{I_\nu}^2   \int {d^3k \over (2\pi)^3} P_{\rm HI}(k)\nonumber \\
&\times &\int d^2\theta  \exp\left(i2\pi\vec\theta\cdot\left({\bf u}_\nu-\frac{{\bf k_\perp}r_\nu}{2\pi}\right) \right) A_\nu(\vec\theta) \nonumber \\
& \times & \int d^2\theta' \exp\left(- i2\pi\vec\theta'\cdot\left({\bf u'}_{\nu}-\frac{{\bf k_\perp}r_{\nu}}{2\pi}\right)\right) A_\nu(\vec\theta')\nonumber \\
 &\times &\int dx \exp\left[i2x \left(2\pi\tau+k_\parallel dr_\nu/d\nu \right) \right ]
\label{etacorrela2}
\end{eqnarray}
The integral over $x$ can readily be carried out now. The dominant 
contribution to the integral comes from $\tau \simeq k_\parallel/(2\pi) dr_\nu/d\nu$, which establishes the correlation scale in the direction along the line 
of sight. The  variation of  frequency dependence of integrals
over $\theta$ and $\theta'$  is expected to be small for the bandwidth
of MWA and therefore these integrals can be computed at some frequency
that lies within the bandwidth.     In this
paper we assume   these values to be  fixed at the central frequency $\nu = 154 \, \rm MHz$ and use them to compute the relations in Eq~(\ref{ukrelations}).  If the 
frequency dependence of the primary beam and the background intensity 
are neglected, the integral over $\nu$ in trivial. As noted above, this is a good  approximation for  MWA.  The power spectrum of the HI signal, based on Eq.~(\ref{etacorrela2}), is shown in Figure~\ref{signal_ps}. In this calculation, we use the HI power spectrum $P_{\rm HI}(k)$ given by the simulation of \citet{Furlanetto et al}.

Eqs.~(\ref{etacorrela2}) and~(\ref{viscorrfreq}) can be used to infer
many important properties of the HI signal: 
\begin{itemize}
\item[1.] There exists a near one-to-one relation between {\bf u}, $\tau$ 
and the Fourier components of the HI power spectrum (\citet{Parsons 2012a, Delay 1, Drift paper, Morales Hewitt}):
\begin{equation}
 u_\nu=\frac{k_{\perp 1} r_\nu}{2\pi},\hspace{10 pt} v_\nu=\frac{k_{\perp 2} r_\nu}{2\pi},\hspace{10 pt} \tau=\frac{k_\parallel c(1+z)^2}{2\pi H_0 \nu_{21}E(z)} 
\label{ukrelations}
\end{equation}
where $\nu_{21}$ is the rest frame frequency of the 21 cm line, $r_\nu$ is the transverse comoving distance, $E(z)=\sqrt{\Omega_M (1+z)^3+\Omega_k(1+z)^2+\Omega_\Lambda}$ and z is the redshift corresponding to the observed frequency $\nu$.
The relation between $k_\parallel$ and $\tau$ follows from the relation: 
$2\pi\tau = k_\parallel dr_\nu/d\nu$ (Eq.~(\ref{etacorrela2})). As noted above,
all the frequency dependent quantities in Eq.~(\ref{ukrelations}) are computed
as a fixed frequency $\nu_0 = 154 \, \rm MHz$. 
\item[2.] The correlations in the sky plane and along the line of sight
are nearly separable. This allows us to compute weights in the plane of the 
sky owing to  $w$-term  and the distortion of intensity pattern in a tracking run (the next two subsections) without the 
additional complication owing to frequency dependence of these quantities. 
\item[3.] Eqs.~(\ref{etacorrela2}) and~(\ref{ukrelations})  allow us to simplify the relation between visibility correlation and the HI power spectrum. Eq.~(\ref{etacorrela2}) can be solved in the limit defined by Eq.~(\ref{ukrelations}) 
to give (e.g. \citet{Delay 4, Pen et al 2009, Morales 2005, McQuinn et al 2006}):
\begin{equation}
\langle V_\tau({\bf u}) V_\tau ^* ({\bf u}) \rangle = {\bar{I_\nu}^2 \Delta\nu \theta_0^2  \over r_\nu^2 dr_\nu/d\nu} P_{\rm HI} (k)
\label{normali}
\end{equation}
Here the MWA primary beam solid angle  $\theta_0^2 = \lambda^2/A_{\rm eff}$. 
For MWA $A_{\rm eff} = 21.5 \, \rm m^2$ at 150 MHz (\citet{MWA}). $\Delta\nu = 10.2 \, \rm MHz$ is the 
total band width we use in this work.  The mean specific intensity ${\bar{I_\nu}} = 2k_BT_B/\lambda^2$. This allows us to express the HI signal as the square of the  product 
of mean brightness temperature and the HI power spectrum $T_B^2 P_{\rm HI}(k) $  in the units $(mK)^2 (Mpc/h)^3$. It should be emphasized that Eq.~(\ref{normali})  provides the suitable normalization only when ${\bf u'} = {\bf u}$, $w=0$, 
as has been assumed throughout this section, and the impact of sky intensity distortion while  tracking a region
is not considered. All these effects act to  lower  the RHS of Eq.~(\ref{normali}),  the measured visibility correlation,  for a fixed signal $T_B^2 P_{\rm HI}(k) $. This is accounted for by appropriate weights we discuss in the 
next two sections. 

\end{itemize}
\subsection{HI signal and w-term}

\label{sec:h1sigwterm}

From Eqs.~(\ref{etacorrela1}) and~(\ref{etacorrela2}), we can gauge the  impact of the $w$-term.
These equations and the discussion following them  shows that the angular 
integrals depend only weakly on  the line of sight variables.
 The main effect of the $w$-term is to alter the 
integrals over angles which we study here.

 For a given baseline ${\bf b}$:
$w_\nu = {\bf b} \cdot {\bf s}_0 \nu/c$, where ${\bf s}_0$ is the phase center at 
any time. As a region is tracked, the $w$-term changes owing to the drift of the phase center. For a tracking run, ${\bf u}_\nu^2 + w_\nu^2$ is left invariant at
any frequency; this result simply follows from the fact the the baseline 
length is fixed. 

After the inclusion of the $w$-term, the measured visibility for a given intensity distribution is given as:
\begin{eqnarray}
 && V_\nu({\bf u}_\nu,w_\nu)=\int A_\nu (l,m) \Delta I_\nu(l,m)\nonumber \\
 &\times &\exp{\left[-i2\pi\left\{u_\nu l+v_\nu m-w_\nu/2(l^2+m^2)\right\}\right]} dldm
 \label{visw}
\end{eqnarray}
Here we have replaced $\vec{\theta}$ with its components (l,m) and also made the approximation:$\sqrt{1-l^2-m^2}-1\simeq -(l^2+m^2)/2$, which is valid for 
MWA primary beam.
After substituting Eq.~(\ref{deli_a}) into Eq.~(\ref{visw}) gives us: 
\begin{eqnarray}
 && V_\nu({\bf u}_\nu,w_\nu)=\bar{I}_\nu\int\frac{d^3k}{(2\pi)^3}\delta_{\rm HI}({\vec {k}})e^{ir_\nu k_\parallel}
 \int  A_\nu(l,m) \nonumber \\
 && \times\exp\biggl[-2\pi i \biggl \{\left(u_\nu-\frac{{k_{\perp 1}}r_\nu}{2\pi}\right)l+\left(v_\nu-\frac{{k_{\perp 2}}r_\nu}{2\pi}\right)m\nonumber \\
 && -  {w_\nu\over2}(l^2+m^2) \biggl \}\biggl]dldm 
\label{viswithw}
\end{eqnarray}
Each MWA tile being approximately a square aperture, the primary beam $A_\nu(l,m)$ can be written as:
\begin{equation}
 A_\nu(l,m)=\frac{\sin(\pi L_x l)}{\pi L_x l}\frac{\sin(\pi L_y m)}{\pi L_y m} \label{pribeam}
 \end{equation}
Here $L_x$ and $L_y$ are dimensionless. They correspond  to 
the ratio of the length of the tile along x- and y-axis  to the wavelength. For central
wavelength of the observation  $L_x = L_y \simeq  2$. Eqs~(\ref{visw}) and~(\ref{pribeam}) show
that integrals over $l$ and $m$ are separable and identical. These integrals 
cannot be done analytically but under certain approximations meaningful 
analytic expressions can be found. Let us define:
\begin{eqnarray}
Q_\nu(k_{\perp 1};u_\nu,w_\nu)  &=& \int dl \exp\biggl [ -2\pi i \biggl \{\left(u_\nu-\frac{{k_{\perp 1}}r_\nu}{2\pi}\right)l \nonumber \\
&& - {w_\nu\over2}l^2 \biggl \} \biggl ] \frac{\sin(\pi L_x l)}{\pi L_x l}
\label{intprib}
\end{eqnarray}
$Q_\nu(k_{\perp 1};u_\nu,w_\nu)$ is a function of $k_{\perp 1}$ and is parametrized by
$u_\nu$ and $w_\nu$. First we consider, $w=0$. In this case, it can be shown that 
if the limits of the integral are  allowed to go from minus infinity to plus infinity, we obtain, 
\begin{eqnarray} 
Q_\nu(k_{\perp 1};u_\nu,0) & = &  {1 \over L_x} \, \, \>  {\rm if} \, \,  |(u_\nu - r_\nu k_{\perp 1}/(2\pi))|/L_x < 1/2  \nonumber \\
& = & 0 \, \, \> \hbox{otherwise}
\end{eqnarray}
We notice that the approximation used is good  because the function
has a compact support provided by the primary beam. As $L_x \simeq 1/\theta_0$ 
where $\theta_0$ is the extent of the primary beam, this result means that, 
for a given $u_\nu$, the wavenumbers that contribute to the integral are the ones
that are bounded by the extent of the primary beam. This result is already
implied by  Eq.~(\ref{etacorrela1}). 

\begin{figure}
\centering
\includegraphics[width=0.5\textwidth]{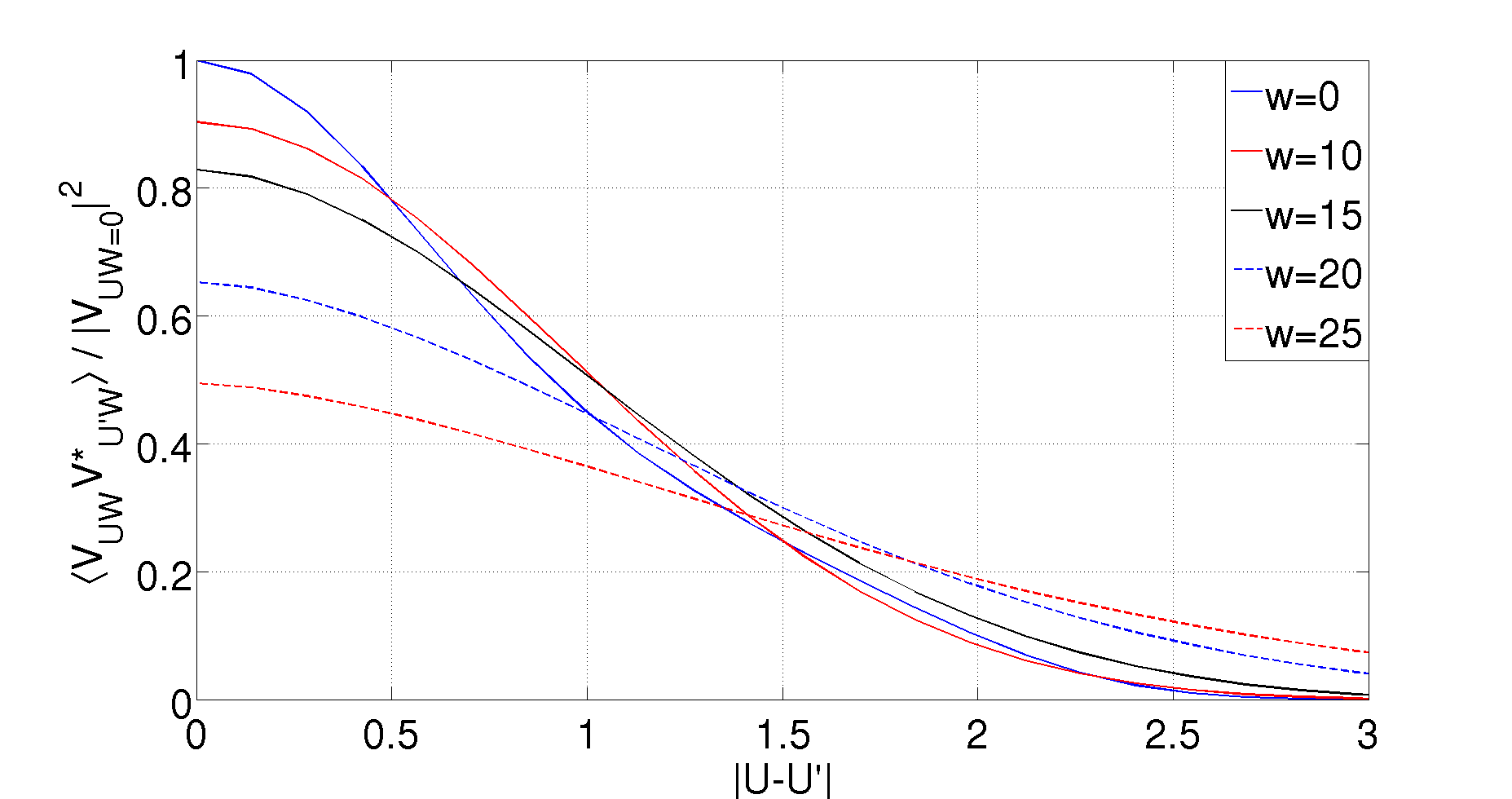}
\caption{The HI signal is plotted as a function $|{\bf u} -{\bf u'}|$ for 
different values of w. The impact of w-term is to decrease the overall
signal owing to shrinking of the primary beam and increase the correlation
length $|{\bf u} -{\bf u'}|$.}
\label{fig:wterm1}
\end{figure}

Eq.~(\ref{intprib}) cannot
be analytically approximated so readily for non-zero $w$. We  use the 
stationary phase approximation to analytically evaluate the integral. For this assumption
to hold, the phase of the exponent  should be much larger  than the 
slow variation of the primary beam. This would be the case if $wl^2$ is large. 
In this approximation, we obtain:
\begin{eqnarray}
Q_\nu(k_{\perp 1};u_\nu, && w_\nu) = \sqrt{{2i\over w}}{\sin(\pi L_x (u_\nu-k_{\perp1} r_\nu/(2\pi))/w_\nu) \over \pi L_x (u-k_{\perp 1} r_\nu/(2\pi))/w_\nu} \nonumber \\
&& \times\exp\left [-i(u_\nu-k_{\perp 1} r_\nu/(2\pi))^2/(4\pi w_\nu)\right ]
\label{wtermpb}
\end{eqnarray}
The main impact of the inclusion of the $w$-term can be discerned from this expression. In the 
limit of large $w$, the impact  of the $w$-term is to shrink the MWA beam and the primary beam  tends to $1/\sqrt{w}$ (\citet{W projection}). \footnote{The 
impact of $w$-term can be more readily computed if the beam is Gaussian (e.g. 
Appendix B of \citet{Drift paper}) In this case, the primary beam  approaches
$(1/\theta_0^2+\pi w_\nu)^{-1/2}$ for non-zero $w$.}  However, this also means that the 
spread of $k_{\perp 1}$ for which the integral is non-zero also increases, as
seen in the terms involving the  $sin$ function. If the decrease of primary beam
results in a loss of signal-to-noise, an increase in the  correlation length $|{\bf u} -{\bf u'}|$
gains  signal-to-noise. We can write visibility correlation  for pairs of ${\bf u}, w$ and ${\bf u'}_\nu, w_\nu'$ as:
\begin{eqnarray}
&&\langle V_\nu({\bf u}_\nu, w_\nu) V_\nu ^* ({\bf u'}_\nu, w_\nu') \rangle = \int {d^3k \over 2\pi^3} P_{\rm H1}(k)  Q(k_{\perp 1}; u_\nu, w_\nu)\nonumber \\
&&\times Q(k_{\perp 2}; v_\nu, w_\nu) Q(k_{\perp 1}; u_\nu', w_\nu') Q(k_{\perp 2}; u_\nu', w_\nu')
\label{wtermvisvis}
\end{eqnarray}
Eq.~(\ref{wtermvisvis}) can be computed numerically. In Figure~\ref{fig:wterm1}, we show how the HI correlation function is affected in the presence of 
$w$-term. 
These expressions are also valid for diffuse foregrounds which  have 
 a different 2-dimensional power spectrum and frequency dependence, e.g. 
 optically thin Synchrotron radiation for which  the angular and frequency dependence is  separable. 
As Eq.~(\ref{wtermvisvis}) can be used to compute the impact of $w$-term at 
any frequency, it can readily be generalized to study diffuse foregrounds.

\subsection{Time dependent coordinate system and w-term}
\label{sec:movinggrid}

In a tracking  interferometric observation, a phase 
center is tracked and snapshots are taken at regular intervals with 
short duration. Each of these snapshots can be imaged and the images 
added if the successive   fields of view can be  assumed to be coplanar.
 This approximation breaks down for wide field-of-view instruments such as MWA. One manifestation of the wide field-of-view is the $w$-term whose
impact was studied in the previous sub-section. In this section we generalize the discussion of the last sub-section to take into account the time 
dependence of the  non-coplanarity of the tracked region (\citet{1999ASPC..180.....T}).

As the region is tracked,  the relation between the image and 
astronomical  coordinates changes which distorts the intensity 
pattern with respect to the phase center being tracked. More specifically, this effect arises from the  projection of a non-coplanar array on a plane, 
which is necessary to perform the Fourier transform for imaging. It is best 
illustrated  with a set of point sources. These sources appear to move with respect to the phase center (e.g. Fig. 19-9 in \citet{1999ASPC..180.....T}). 
The distortion of the intensity pattern corresponds to non-uniform 
stretching and  it increases  for sources further away
from the phase center. Thus this effect can not be corrected  by a standard shift of coordinate.  The non-uniform stretching makes the situation complex, and the standard grid approach is difficult to implement in this case. For 
a set of point sources, the correction for  this effect could   be applied
iteratively in the image plane (Chapter 19, \citet{1999ASPC..180.....T}). 

For a small field-of-view, this effect can be neglected
and a unique coordinate system (e.g. time independent direction cosines $\{l,m\}$) can be used to relate 
the image coordinates with the sky intensity pattern for a long tracking run.
However, it is not possible to define such a coordinate system when either 
the field of view is large or the tracking period is long. 

Our aim here is not to correct for this effect but rather to estimate its impact on
the correlation of visibilities at two different times during a tracking run: 
suppose we measure visibilities within a  small cell in the $u$--$v$ plane (the size of the cell  will be discussed in a later section)  centered around  a baseline $\{u,v\}$ at 
$t=0$. At a later time $t = t'$ another baseline might enter this cell. From the discussion in the previous subsections (e.g. Figure~\ref{fig:wterm1}) the two 
 visibilities are expected to correlate strongly with each other  (even if the values of 
$w$  differ significantly for these two sets  this statement is generally true). However, visibilities measured at two different times do not 
correspond to the same  intensity pattern. Our aim here is to estimate  the level of de-correlation caused 
by the distortion of intensity pattern during a tracking run. In this paper, we construct a time-dependent coordinate system  which allows
us to analyse this distortion of intensity pattern. We assess the impact of this 
effect when a region   is tracked using the MWA primary beam. In particular, 
we consider this effect on the visibilities produced by the EoR HI signal.

We start by recalling the definition of  direction cosines for a point on the sky whose coordinates, declination and hour angle $\{\delta,h\}$, are:
written as (\citet{Chris}):
\begin{eqnarray}
l & = & \sin(h)\cos(\delta) \\
m & = & \cos(\delta) \cos(h) \sin(\phi) -\sin(\delta)\cos(\phi) \\ 
n & = & \cos(\delta) \cos(h) \cos(\phi) +\sin(\delta)\sin(\phi)
\end{eqnarray}
It can be shown that $n = \sqrt{1-l^2-m^2}$. The phase center is 
always defined as $l = 0$, $m = 0$, $n=1$; for the coordinates defined
above it is: $h =0$ and $\delta = \phi$. 

As a phase center is tracked  owing to the rotation of the Earth, $\delta$ 
remains fixed but the hour angle changes. For a wide field of view, 
this can result in distortion of the intensity pattern of the sky. 
To take into account this effect, we can define a time-dependent coordinate
system:
\begin{equation}
l(\delta h)= \sin(h+\delta h)\cos(\delta) - \sin(h_0+\delta h)\cos(\delta_0)
\end{equation}
Here $h_0$ and $\delta_0$ define the phase center for $\delta h = 0 $; $\delta h$
defines the flow of time. $m$ can be similarly defined and $n$ can be computed from $l$ and $m$. This definition gives  a time dependent 
coordinate system where the coordinates are  always defined with respect to the 
phase center. It is easy to verify that for small field of view 
and for small tracking times,  which corresponds to  cases when higher order terms in $h$, $\delta h$ and $\delta$ can be dropped, $l$ is independent of 
time which means that the distance of a point from the phase center is left
invariant under tracking. In such cases, the intensity pattern on the sky 
corresponds to the unique  intensity pattern defined by sky coordinates 
$\delta$ and $h$ and  remains unchanged  as the phase center is tracked. 

However, when this approximation breaks down, $l$ becomes a function of 
time and it is impossible to define a unique relation between  direction
cosines and sky coordinates. This means that any quantities defined 
with respect sky coordinate (e.g. intensity pattern) become time dependent.
The visibility for the HI signal is given by:
\begin{eqnarray}
&& V_\nu({\bf u}_\nu, w_\nu,\delta h)=\bar{I}_\nu\int\frac{d^3k}{(2\pi)^3}\delta_{\rm HI}({\vec {k}})e^{ir_\nu k_\parallel}
 \int  A_\nu(l,m) \nonumber \\
 && \times\exp\biggl[-2\pi i \biggl \{\left(u_\nu-\frac{{k_{\perp 1}}r_\nu}{2\pi}\right)l+\left(v_\nu-\frac{{k_{\perp 2}}r_\nu}{2\pi}\right)m \nonumber \\
 && -  {w_\nu\over2}(l^2+m^2) \biggl \}\biggl]\sin\delta d\delta dh
\label{viswithw}
\end{eqnarray}
The direction cosines $l$ and $m$  are now  functions of time. The angular 
integral is carried out  over $\delta$ and $h$. Unlike the earlier case (fixed
grid) this is not a product of two one-dimensional integrals. The correlation
of the visibilities $V_\nu({\bf u}_\nu, w_\nu,\delta h)$ can be computed using the 
same methods as outlined in the previous sections. 

\begin{figure}
\centering
\includegraphics[width=0.5\textwidth]{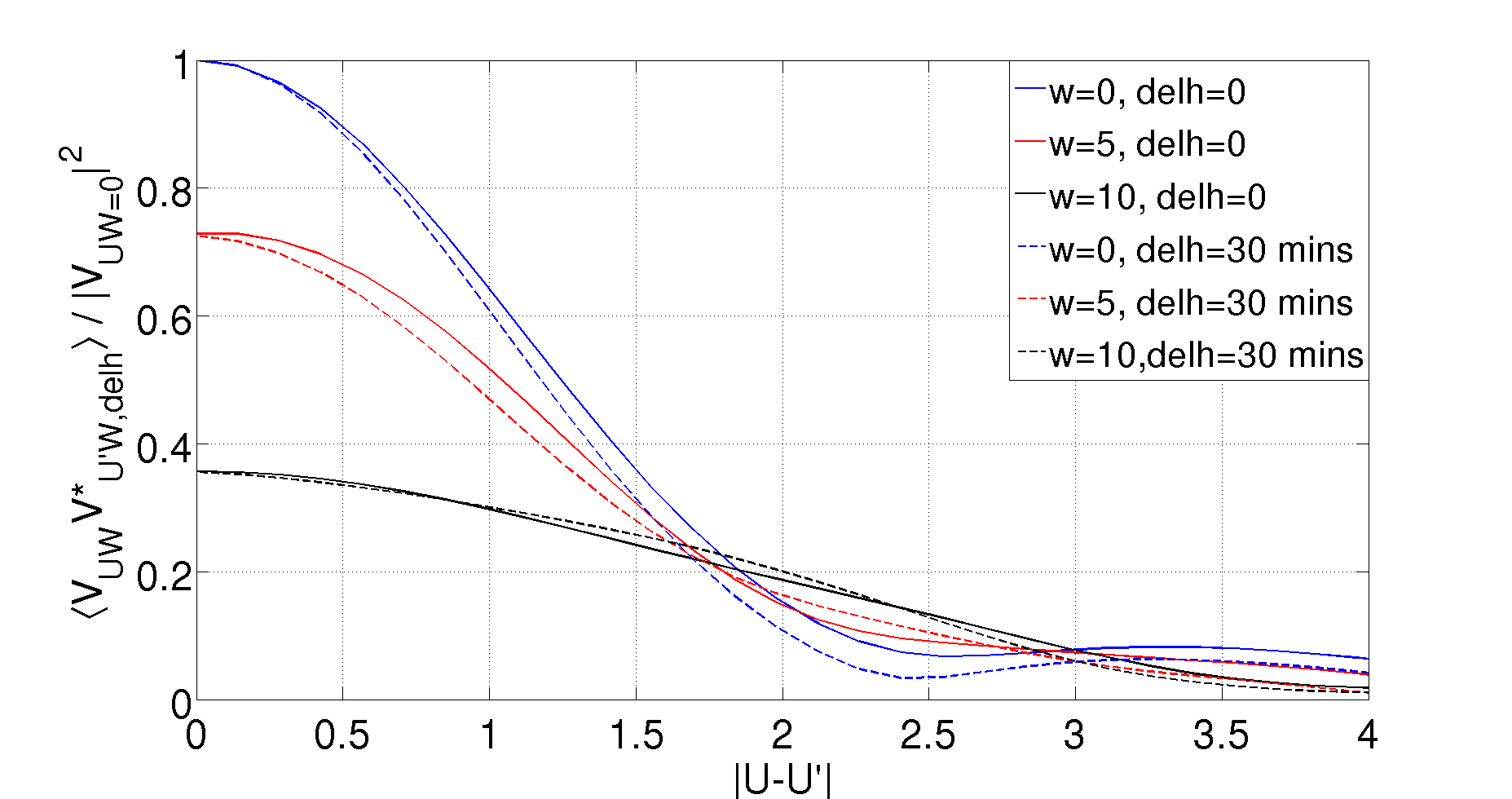}
\caption{The HI signal is plotted as a function $|{\bf u} -{\bf u'}|$ for 
different values of w and $\delta h$. The initial 
phase center is chosen  to be $h=0$ and $\delta = \phi$. }
\label{wterm2}
\end{figure}

In Figure~\ref{wterm2} we show the results when the effect of 
the time dependent coordinate system ('moving grid')
is included.  The initial 
phase center ($\delta h =0$) is chosen  to be $h=0$ and $\delta = \phi$.
The results are shown for two different values of $\delta h$ and a range of 
$w$ values. We only show the auto-correlation function for a given value of 
$\delta h$. But the results shown in Figure~\ref{wterm2} can be used
to assess the  cross-correlation of  visibilities measured at two different times.  For our case the  value of this cross-correlation 
 lies between the auto-correlations of visibilities  measured at the same time. The moving 
grid doesn't introduce another scale in the problem and the results
in this case are not very different from the case for a fixed grid. In both
cases the dominant correlations occur for  ${\bf u} - {\bf u'} < 1.5$. 

Figures~\ref{fig:wterm1} and~\ref{wterm2} are based on MWA primary beam. However, it is possible to glean generic information applicable for other primary beams from them.  First,  the decorrelation length $|{\bf u}-{\bf u'}|$ scales as the inverse of the primary beam (e.g. \citet{Drift paper}). So for a smaller beam, the decorrelation seen in the Figures as a function of $|{\bf u}-{\bf u'}|$  would be shallower. The impact of the $w$-term for a different primary beam can be partially gauged from Eq.~(\ref{wtermpb}), which is valid for large values of $w$. 
In this limit, the primary beam tends to $1/\sqrt{w}$, irrespective of the primary beam of the telescope. It is difficult to analytically estimate the impact of the $w$-term when this limit does not hold. But it can be shown that the impact of $w$-term diminishes for a smaller primary beam, e.g. a Gaussian beam for 
which the primary beam tends to $(1/\theta_0^2+\pi w_\nu)^{-1/2}$ for non-zero $w$
(e.g. \citet{Drift paper}). As noted above, the distortion of intensity pattern during a tracking run is a wide field effect. For a smaller primary beam, the level of decorrelation seen in Figure~\ref{wterm2} would be  smaller but it is difficult to analytically estimate it.

\subsection{Weights for cross-correlation}

\label{sec:weightscross}
Eq.~(\ref{viswithw}) can be used to compute the counterpart of Eq.~(\ref{etacorrela1})  which takes into account the impact of non-zero $w$-term and the distortion of intensity pattern. We compute this expression for the 
visibility correlation in delay space numerically. In this formulation, 
the measured visibility is a function of five parameters: $\tau, {\bf u}, w, t$.
Here, as noted above, ${\bf u}$ and $w$ are the values of these variables
at a fixed frequency which we choose to be $\nu_0 = 154 \, \rm MHz$.

We define the weight on a given cross-correlation as:
\begin{equation}
{\cal W}({\bf u},w,t;{\bf u'},w',t') = {\langle V_\tau({\bf u},w,t) V_\tau ^*({\bf u'},w',t') \rangle  \over \langle V_\tau({\bf u},0,t) V_\tau ^*({\bf u'},0,t') \rangle }
\label{weightsdef}
\end{equation}
The weights are defined with respect to the HI cross-correlation 
computed  in Eq.~(\ref{etacorrela1}) for ${\bf u} = {\bf u'}$, $w = w' = 0$,
$t = t'$ and $\tau = \tau'$. We only consider the case $\tau' = \tau$ for the 
computation of weights. 

 Using Eq.~(\ref{weightsdef}) allows one to recover the HI power spectrum 
for a fixed wave number $k$ from visibility cross-correlations.

\section{Analysis of MWA data}

The MWA is a low frequency radio interferometer array located in  Western Australia. It consists of 128 antenna tiles with each tile comprising of 16 crossed dipole antennas over a metal ground screen in 4 x 4 grid. MWA bandwidth  is 30.72 MHz, divided into 24 coarse channels of width 1.28 MHz each. The total bandwidth is  divided into 768 fine channels. With the use of an analog beamformer appropriate phase delays are introduced in each individual dipole antenna to track the pointing center of the beam across the sky. For more information on MWA please see \citet{MWA, Lonsdale et al}. 

\begin{figure*}
\centering
\includegraphics[width=0.9\textwidth]{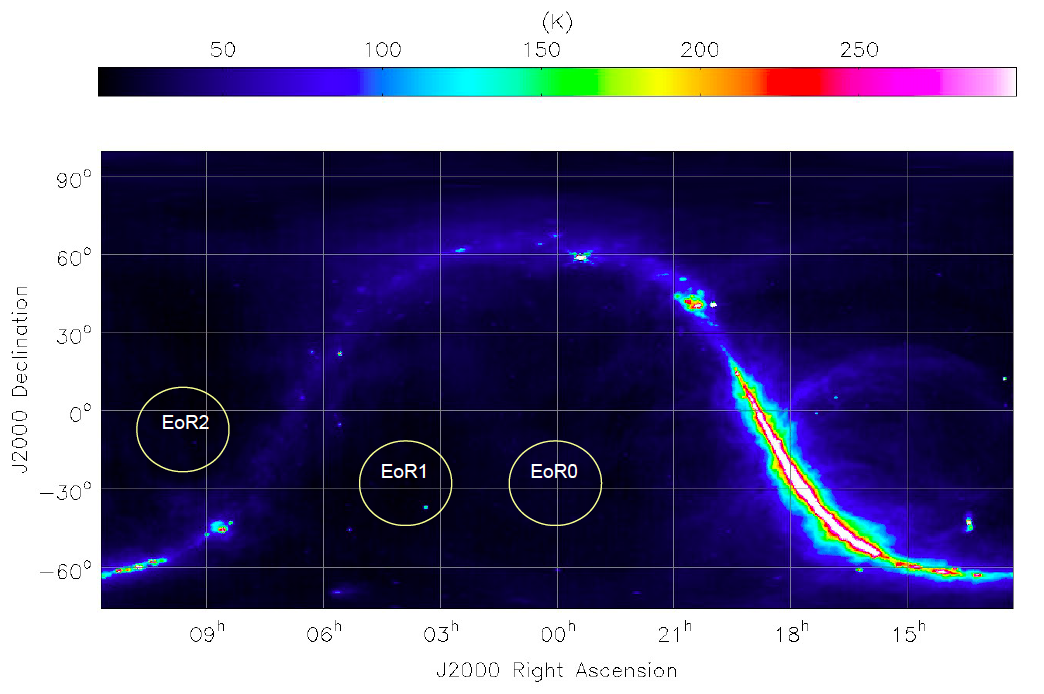}
\caption{EoR fields are shown on the Galactic map at 408 MHz (\citet{Haslam map}). Three fields, away from the galactic plane, have been identified for the MWA reionization study. These are EoR0, EoR1 \& EoR2 respectively as shown in circles. In this paper we focus on the EoR1 field centered at RA $=$ 4h, Dec $= -27 ^\circ$.}  
\label{eor_fields}
\end{figure*}

To minimize the effect of Galactic synchrotron emission, the MWA EoR community has chosen three fields on the sky away from the Galactic plane. These fields have been named as EoR0, EoR1 and EoR2 and  are  shown in Figure~\ref{eor_fields}. In this paper we present 3~hours of tracking analysis of the 
EoR1 field centered at RA $=$ 4h, Dec $= -27 ^\circ$.

Many  research groups are currently developing pipelines to extract statistical information from radio interferometric data, with an aim to detect the HI 
signal from EoR (\citet{Jacobs et al, Hazelton et al, Dillon et al, Trott et al}). These can be divided broadly into two categories: image based and  visibility based pipelines. 
For foreground subtraction and  imaging  these pipelines use  the following imaging algorithms: Real Time System (RTS; \citet{Mitchell et al, Ord et al}) and  Fast Holographic Deconvolution (FHD, \citet{Sullivan et al}).

The image based pipelines (\citet{Dillon et al, Hazelton et al}) use source catalog created through the deconvolution of the data which is subtracted to obtain a residual image cube. The Fourier transform of this image cube with some further processing yields the power spectra. On the other hand, the visibility based pipelines (\citet{Trott et al}) use the data in visibility domain 
 for power spectra estimation, after the initial processing in the 
image domain   for obtaining the foreground model.  The detailed comparison of the outputs from all the methods described above is provided in \citet{Jacobs et al}. \citet{Delay 4, Thyagrajan 2015b} describe the impact of wide field of view in power spectra estimation. 

A special variant among the visibility based estimators is `delay spectrum' (\citet{Pober PAPER, Delay 1, PAPER}), which directly Fourier transforms each calibrated visibility along its frequency channels. The radio interferometer PAPER uses this approach extensively; it relies upon redundant baselines to calibrate the interferometer and  East-West and near East-West baselines for power spectrum estimation (\citet{Delay 1, PAPER, Ali et al PAPER}). This particular scheme has been discussed and implemented only for  redundant drift scan observations.  In this paper we explore the possibility of applying this approach for non-redundant imaging arrays and tracking measurements.

In this and the next section, we discuss in detail our method of MWA data analysis and power spectrum estimation from the  data.

 We summarize below the major ingredients of the  method
and then describe each of the stpng in detail in subsequent sections:
\begin{itemize}
\item[1.]  CASA (\citet{CASA}) is used for initial processing of the data to calibrate raw visibility measurements. This 
is followed by the creation of a model sky image from clean components. This 
model is then subtracted in the visibility domain to obtain residual visibilities. We use both the calibrated and residual  visibilities for computing the 
power spectrum. 
\item[2.]   Each visibility is then Fourier transformed in frequency space (Eq.~(\ref{visetaspace})). This process is needed for isolation of foregrounds
in the  $k_\perp\hbox{--}k_\parallel$ plane. We note the our method 
utilizes both the subtraction of foregrounds and their isolation. But  it does not employ an external point source catalog.
\item[3.] The procedure outlined above yields complex visibilities as a function of five variables: $V_\tau(u,v,w,t)$. For computing the power spectrum we cross-correlate these visibilities for $t'\ne t$ to remove the noise bias. To weigh each cross-correlation we assume that there exist regions in $k_\perp\hbox{--}k_\parallel$ plane which are dominated by only noise and the HI signal.  This allows us to compute 
a weight for each cross-correlation based on the expected HI signal. For computing these weights we take into account the impact of $w$-term and the distortion of intensity pattern  in a tracking scan.  The relevant
method  is elaborated in detail in sections~\ref{sec:h1signal}, \ref{sec:h1sigwterm}, \ref{sec:movinggrid}, and~\ref{sec:weightscross}  and summarized in section~\ref{sec:weights}. 
\item[4.] In section~\ref{sec:powspecest}, we describe the power spectrum estimator,  
taking into account weights given by the expected HI signal, in 3-, 2-
and 1-dimension. We also discuss our method to compute the errors on 
the estimated power spectrum.  
\end{itemize}

\subsection{CASA processing}


 MWA data were  collected at 2-minute intervals with a time resolution of 0.5 seconds and frequency resolution of 40 kHz.  The central frequency of these observations is 154.24 MHz.
For preprocessing we have used the Cotter pipeline (\citet{Offringa et al}) to average to 10 seconds of integration;  we have not performed any averaging over the frequency channels. Cotter also uses the in-built AOFlagger to flag and remove radio frequency interference.  The edge channels of each coarse band are flagged with Cotter due to aliasing effects. After this  preprocessing the 
 Cotter pipeline  delivers the data in the CASA readable `Measurement set (ms)' format for further processing. 

\begin{figure}
\centering
\includegraphics[width=0.5\textwidth]{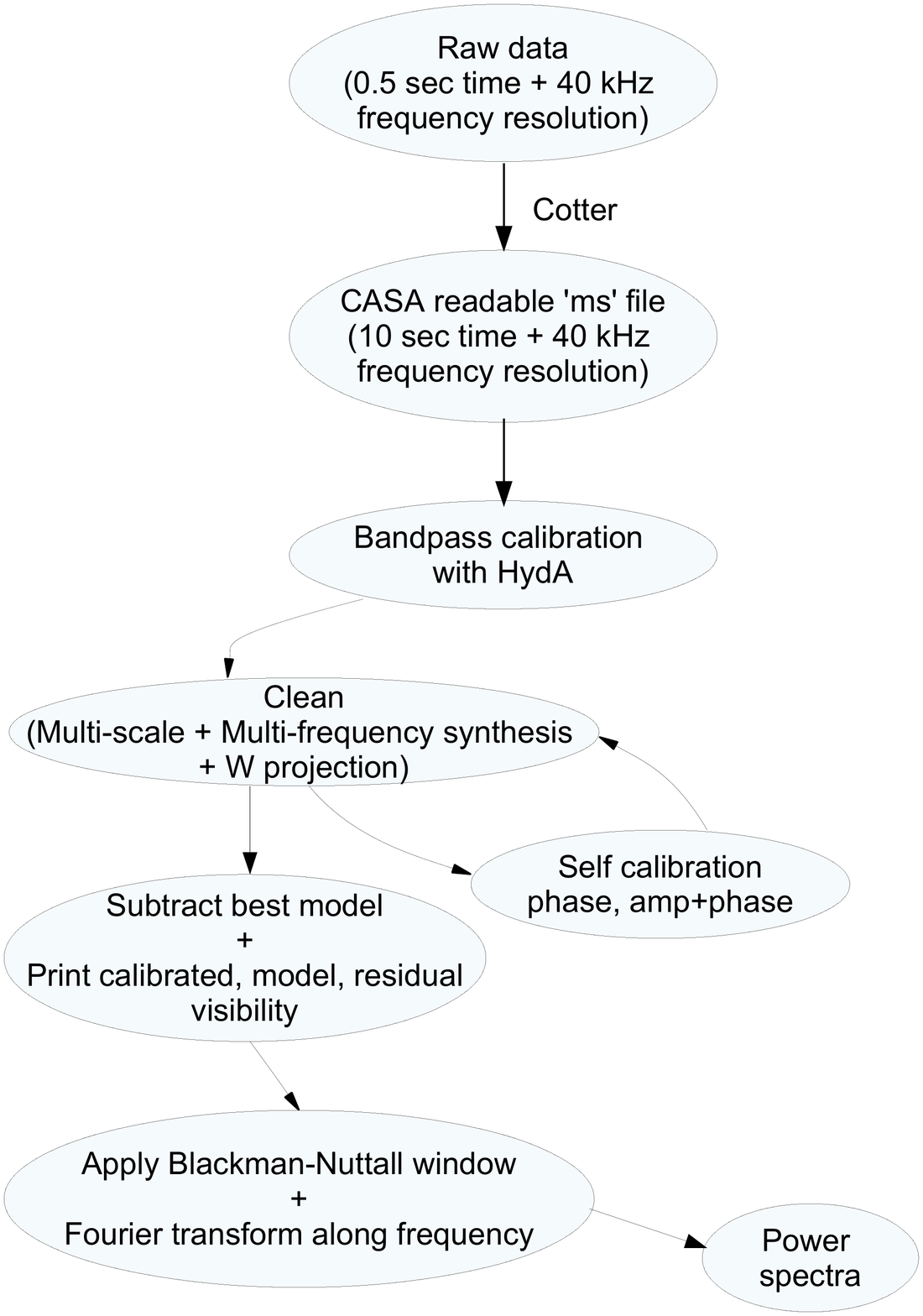}
\caption{A schematic of the  power spectra pipeline is shown.}
\label{pipeline}
\end{figure}

\begin{figure*}
\centering
\subfloat[]{\includegraphics[width=0.9\textwidth]{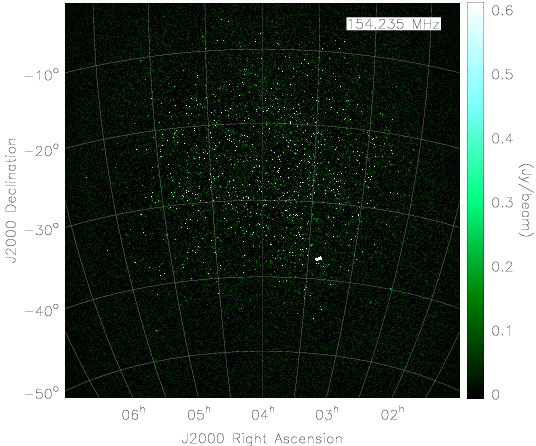}}\\
\subfloat[]{\includegraphics[width=0.5\textwidth]{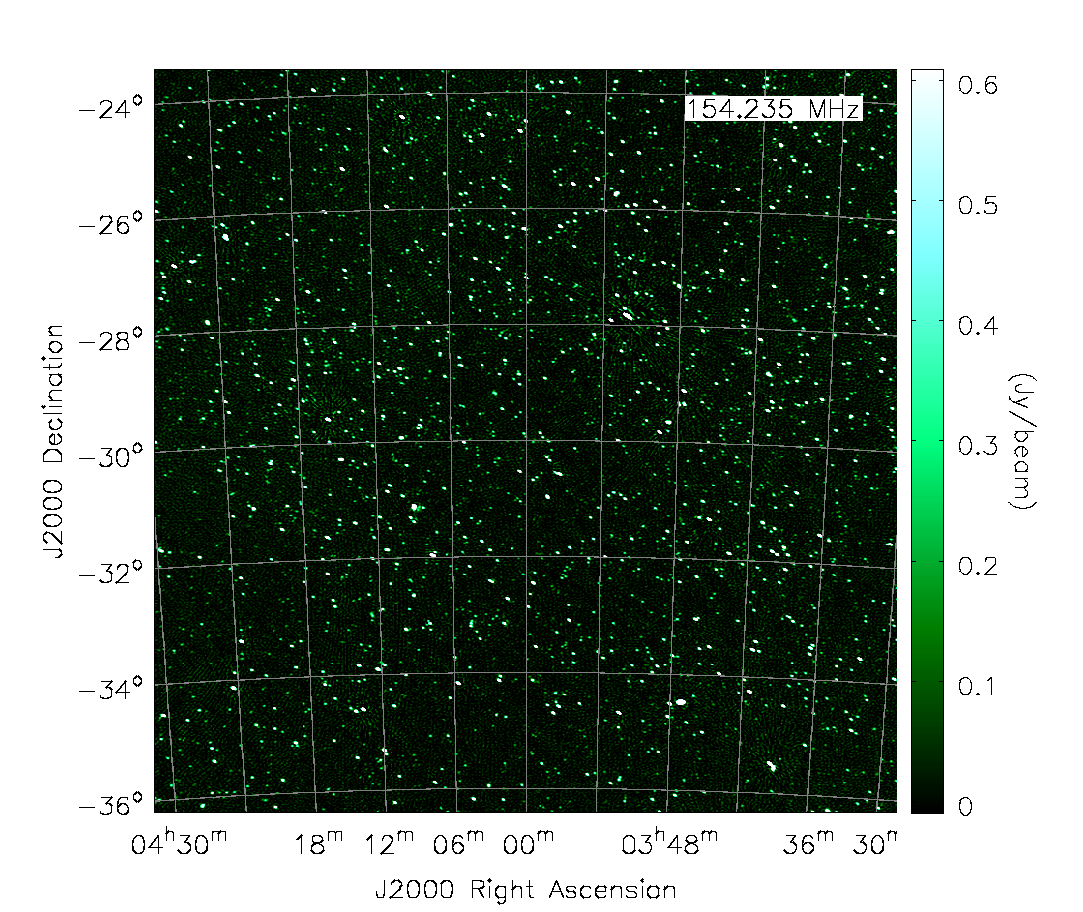}}
\subfloat[]{\includegraphics[width=0.5\textwidth]{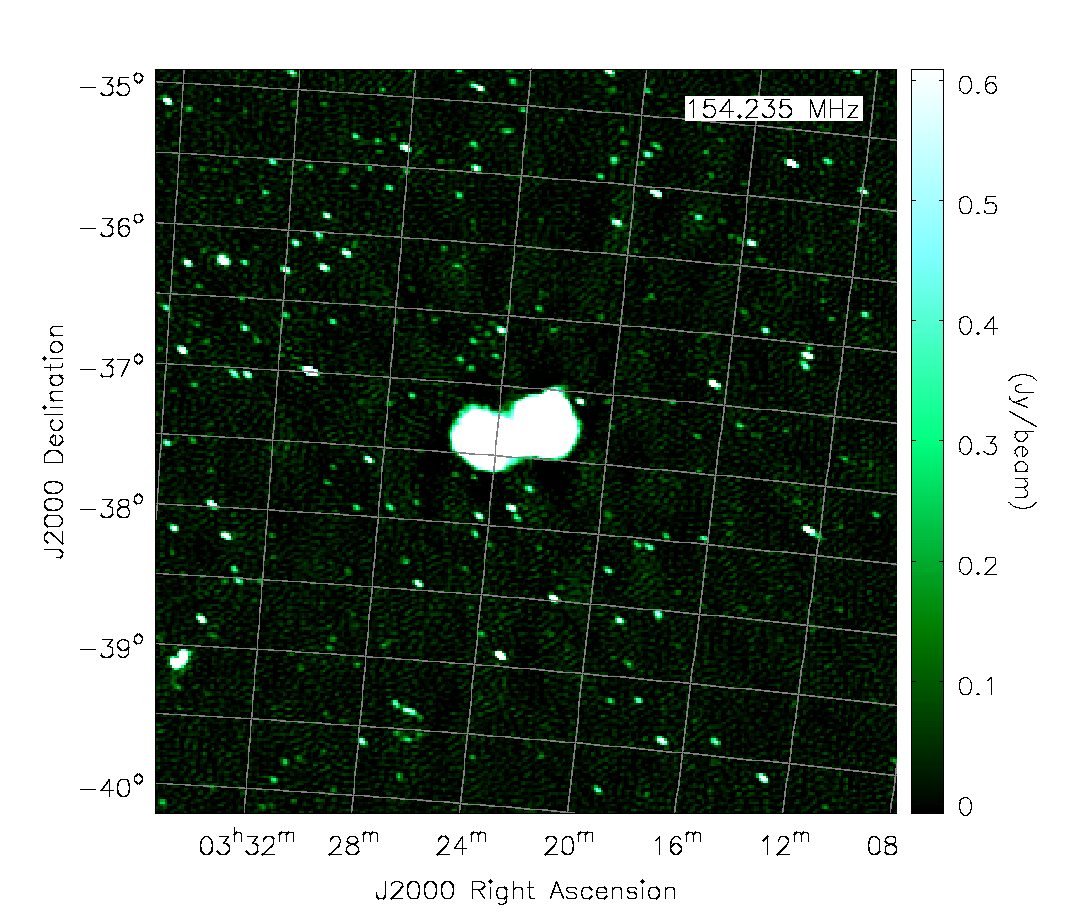}}
\caption{The image of EoR1 field at 154.24 MHz is shown in the top
panel for two minutes of data.  The bottom panels display two regions from the image. 
The   right panel displays  the region containing  the Fornax~A.}
\label{eor1_image}
\end{figure*}

Once the `ms' files are produced for each 2-minute data set, we process each of these 2-minute data in CASA to produce an image. The Hydra A source is used to calculate the bandpass solutions which are applied to the uncalibrated data. We next   construct   a  sky model from these data so that we could subtract it to obtain the residual visibility. After the bandpass calibration the  first round of `clean' is applied on each 2-minute data set. The multi-scale multi-frequency synthesis algorithm (\citet{MSMFS}) has been used for imaging. We have created images of size 3072 x 3072 pixels with 1 arc-minute cell size using the Cotton-Schwab CLEAN (\citet{Schwab}) with uniform weighting scheme. After the first round of clean we have performed self calibration (both phase and amplitude+phase) and apply the  clean loop until the  RMS value of the residual image converges. The threshold limits for the clean stpng were chosen to be 5 sigma. The W-projection algorithm (\citet{W projection}) was also used to correct for the errors arising due to non-coplanarity of baselines.  Once we obtain the best model of the sky for each 2 minute observation, the model visibilities are  then subtracted from the calibrated data using the UVSUB algorithm in CASA to obtain the residual data. This process is followed for both XX \& YY polarizations separately.
 A flow chart of the data pipeline is shown in Figure~\ref{pipeline}. In Figure~\ref{eor1_image}, we present a sample image of  2 minute deconvolution.

As noted above we process the data for only  2 minutes to 
ensure the primary beam doesn't substantially change during the run. 
For a 2-minute scan  we obtain an RMS of 
nearly 40~mJy/beam. 
 
The residual visibility $V_\nu(u_\nu,v_\nu,w_\nu,t)$ is a function of five variables. 
We compute the discrete Fourier transform of the residual visibilities 
in the frequency space weighted by the Blackman-Nuttall (\citet{BN}) window $B_\nu$ to suppress leakage into the EoR window (\citet{Thyagrajan et al, Thyagrajan 2016}):
\begin{equation}
V_\tau(u,v,w,t) = \sum \Delta \exp(i2\pi\nu\tau) V_\nu(u_\nu,v_\nu,w_\nu,t) B_\nu
\label{visetaspace}
\end{equation}
Notice that in  Eq.~(\ref{visetaspace})  the frequency dependence of 
the baseline vector ${\bf b}_\nu =  \{u_\nu,v_\nu,w_\nu\}$ is integrated over. 
Therefore, the labels
 $\{u,v,w\}$ on the LHS of Eq.~(\ref{visetaspace}) need further explanation. 
As noted above (the discussion following Eq.~(\ref{deftau})) 
they    can be chosen to denote a given baseline vector  at a fixed frequency, $\nu_0$. We choose this frequency to be 
the central frequency of the band $\nu_0 = 154 \, \rm MHz$. \citet{Parsons 2012a, Delay 1} provide detail implications of the frequency dependence of the baseline vector.
Here $\Delta = 40 \, \rm kHz$ and 256~channels are used for our study, which
correspond to total bandwidth 10.24~MHz in the frequency range 149.09~MHz to 159.34~MHz.

\section{Power spectrum}

\label{sec:weights}

The visibilities (Eq.~\ref{visetaspace}) are cross-correlated  with 
weights determined from the HI signal (section~\ref{sec:weightscross})  to estimate the power spectrum. For each
pair of parameters, e.g. $\{u,u'\}$, the weights are generally different. 
It is computationally prohibitive to deal with weights for all cross-correlations. We make several simplifying assumptions to make the problem tractable based 
on the properties of the HI signal. In  sections~\ref{sec:h1signal}, \ref{sec:h1sigwterm}, and~\ref{sec:movinggrid} we discuss in detail the HI signal and 
how it is affected by the inclusion of the $w$-term and the additional complication arising from distortion of the field of view as a region is tracked for 
MWA. 

We  summarize the main results of these sections as applied to the data:
\begin{itemize}
\item[1.] In section~\ref{sec:h1signal} the HI signal and its correlations
are discussed in detail. Eq.~(\ref{etacorrela2}) shows that correlations in sky
plane are nearly independent of correlations along the line of sight. This allows us to compute weights for correlations in the plane of the sky independent of 
the third axis. Eq.~(\ref{etacorrela2}) allows us to derive a relation 
between the measured correlation and the inferred HI power spectrum (Eq.~\ref{normali}).  Eq.~(\ref{normali}) defines the scale of cross-correlation weights. 
The weight function ${\cal W}(u,v,w,t; u',v',w',t')$ is unity when 
${\bf u} = {\bf u'}$, $w = w' = 0$ and $t\simeq t'$. Eq.~(\ref{normali}) refers
to this case. 
\item[2.] In section~\ref{sec:h1sigwterm}, the impact of $w$-term on the HI
signal is computed. Eq.~(\ref{wtermpb}) and Figure~\ref{fig:wterm1} capture
the effect of non-zero $w$ on HI correlations. The $w$-term diminishes 
the signal by shrinking the effective primary beam and increases the correlation
length scale $|{\bf u} -{\bf u'}|$. We use the analytic
expression based on Eq.~(\ref{wtermpb}) for computing weights for $w > 30$. 
\item[3.] In section~\ref{sec:movinggrid}, we attempt to assess the 
impact of  time-dependent distortion of intensity pattern in a tracking run
for MWA. Figure~\ref{wterm2} shows the combined effect of moving grid and 
$w$-term. The distortion of intensity pattern  generally acts to enhance decorrelation but is  found to 
be not significant and  doesn't alter the main features
of the signal.  For our computation, we only  update the weights after every 10~minutes to account for this effect. 
\end{itemize}

\subsection{Power spectrum estimator}

\label{sec:powspecest}

As shown above each  correlation receives a different weight depending on the values of $\{u, v,w,t\}$ of the baselines being correlated. As noted above,  we define 
the weights ${\cal W}(u,v,w,t;u',v',w',t')$ such that they approach  unity 
when $u = u'$, $v=v'$, $w = w' = 0$ and $t\simeq t'$ such that the effect of 
the moving grid is not important ($t\ne t'$ for all correlations
to remove noise bias).

 The HI power spectrum is a function of 
$k  = \sqrt{{\bf k}_\perp^2 + k_\parallel^2}$; Eq.~(\ref{ukrelations})
gives the relation between the Fourier components of the HI signal  and 
$\{{\bf u}, \tau\}$. All cross correlations for which the wave vector lies
in some range $k$ and $k+dk$ can be used to construct the unbiased 
HI signal:   $1/N\sum V_\tau(u,v,w,t)V_{\tau'}(u',v',w',t')/{\cal W}$; here $N$ is
the number of   all pairs for which  $k$ lies in the range specified above.   However, this estimator, though unbiased for the HI 
signal, could be dominated by small values of weights ${\cal W}$, which
doesn't make it the lowest noise (or  optimal) estimator.  

As the observed signal is dominated by noise, we consider  an optimal estimator
for our study:

\begin{equation}
\phat{P}_\tau(u,v)={1\over\sigma}\sum_{u',v',w,w',t,t'}\left( V_\tau V_\tau ^* \times {\cal W} \right)
\label{powspecdef}
\end{equation}
where $\sigma=\sum_{u',v',w,w',t,t'} {\cal W}^2(u,v,w,t,u',v',w',t')$.
To avoid noise bias, $t \ne t'$ for 
all cross-correlation. For a given $\{u,v,\tau\}$, Eq.~(\ref{powspecdef}) allows us to compute the 
power spectrum by optimally weighing over all the cross correlations. However, 
as Figures~\ref{fig:wterm1} and~\ref{wterm2} show the correlations
fall substantially for ${\bf u} - {\bf u'} \ge 1.5$ (see also \citet{Drift paper}
and references therein). This motivates us to pixelize the ${\bf u}$-plane
and consider only those visibility pairs  for which the correlations are significant.  We consider cells of different sizes and present results here 
for $\Delta u=\Delta v = 0.5$.  The number of visibility measurements  in a cell vary  depending on the (u,v) values. The shortest baselines  have higher population as expected for  MWA. For 3 hours of analysis and $u, v <  50$, the 
 number of visibilities in a cell lie in the range $\sim 1000\hbox{--}3500$ where each visibility has a time resolution of $\Delta t = 10$ sec. All the cross-correlation within a cell 
are computed using Eq.~(\ref{powspecdef}).

 For averaging over different 
cells,    each cell is assigned an average  weight corresponding to
the RMS of the power spectrum  for a  cell,  $\sigma_p$.  These  weights  
are then used for optimally averaging the power spectrum 
 (Eq.~(\ref{powspecdef}))  over other cells (For details see Appendix~B).  Note that this procedure
allows us to separate  large  correlations of the HI signal, the 
ones for which ${\cal W}$ is close to unity,  from
 the ones which are expected to be  incoherent because ${\cal W}$ is small.

\begin{figure}
\centering
\subfloat[]{\includegraphics[width=0.5\textwidth]{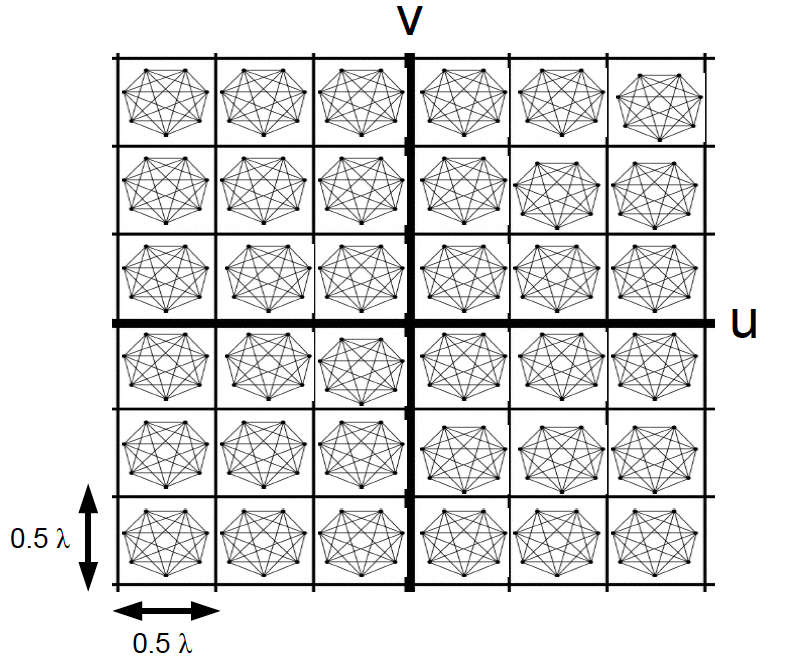}}\\
\subfloat[]{\includegraphics[width=0.5\textwidth]{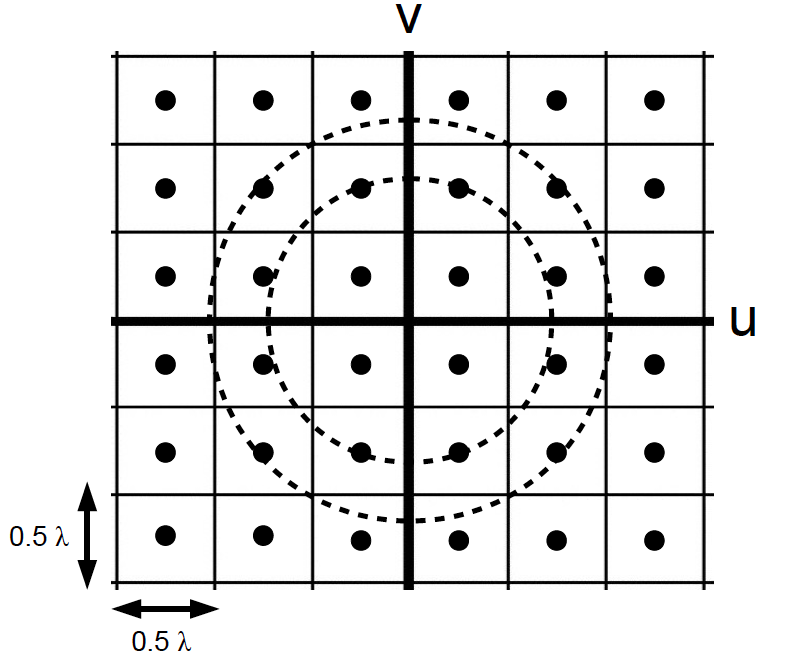}}\\
\caption{(a)This shows the population of visibilities within different uv bins for each $\tau$. The small dots denote visibilities due to individual baselines. These are cross-correlated with each other within a given uv bin. (b) The black dots within a bin are the power value at each bin. An 
optimal azimuthal average, based on weights of each cell,  is done to collapse the uv axes into a single axis of baseline length.}
\label{schem}
\end{figure}

The schematic of the two  processes---the computation of power
spectrum in 3- and 2-dimensions---is displayed  in figure (\ref{schem}):
the top panel  delineates the process of computing cross-correlations
within each cell and the bottom  panel depicts how azimuthal average for 
a fixed baseline length $\sqrt{u^2+v^2}$ is computed. For MWA data, $k_\parallel \gg {\bf k}_\perp$, which means the value of $k$ is dominated by the value of 
$k_\parallel$. This suggests the following method for  computing the 1-dimensional 
power spectrum, which we adopt: all the cells for a given  $\tau$ are  optimally
averaged  using the method described above. This procedure yields a complex 
number. In the Figures that display 2- and 1-d power spectra we plot the absolute value of the estimated power spectrum.

The error on power spectrum in 1-dimension is computed by first estimating 
the RMS for each cell, $\sigma_p$. $\sigma_p$ are then  used as weights for 
 optimal averaging over all the cells
for a fixed $\tau$. The resultant RMS after averaging over the cells approaches $\sigma_{\rm fin} \simeq  (1/\sum\sigma_p^2)^{-1/2}$ if the power spectrum 
across cells is  uncorrelated. This holds for noise but, as noted above,
is an approximation for the HI signal. We expect this assumption to 
be valid in our case as the observed signal is dominated by noise (for detailed
explanation see Appendix~B).

\section{Results}

In Figure~\ref{ps}, the power spectra computed from 3~hours of data are shown in the $k_\perp\hbox{--}k_\parallel$ plane. The power spectra for both  XX and YY polarization are shown for calibrated and residual visibilities. 

\begin{figure*}
\centering
\subfloat[dirty XX]{\includegraphics[width=0.5\textwidth]{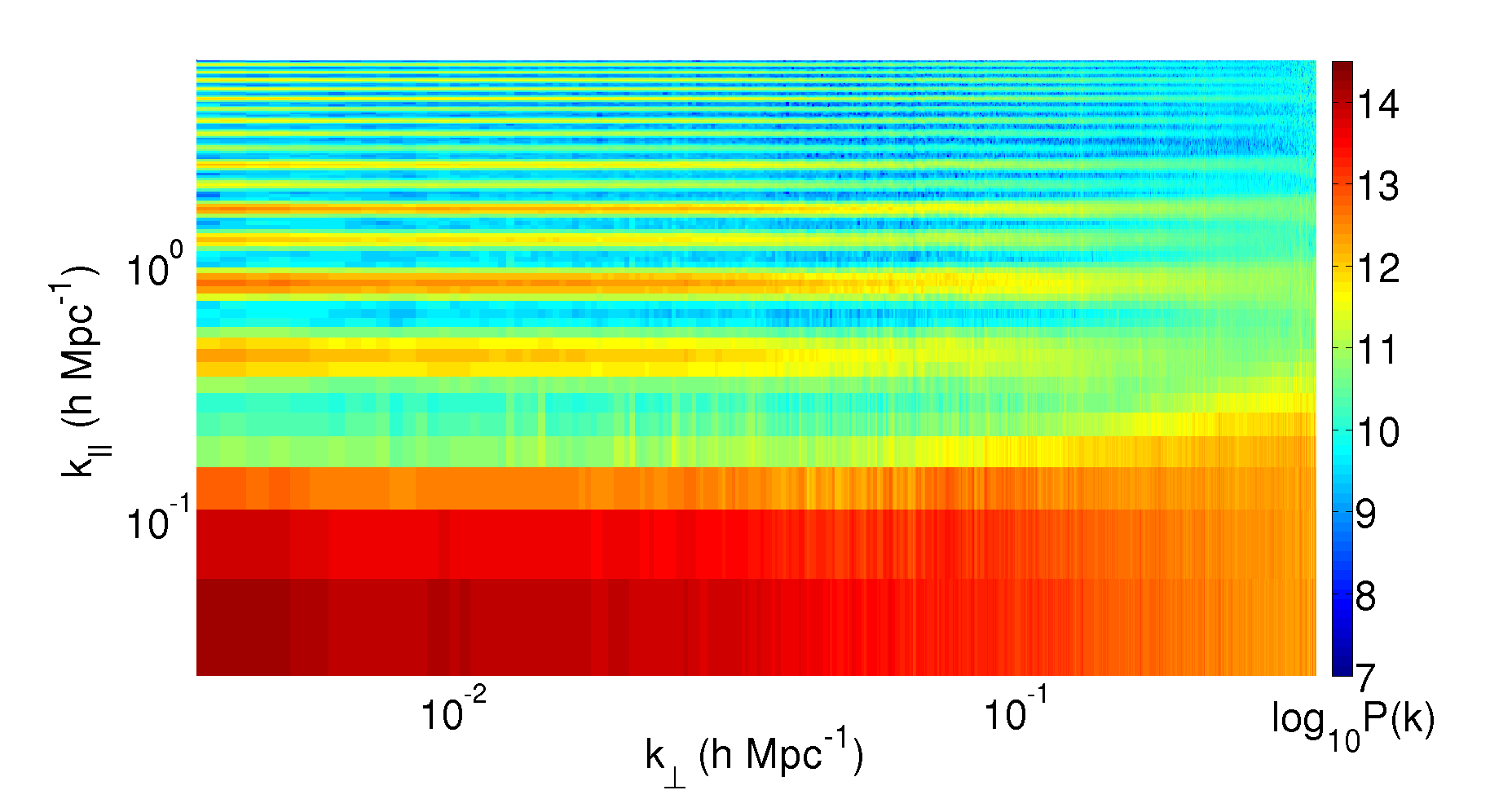}}
\subfloat[dirty YY]{\includegraphics[width=0.5\textwidth]{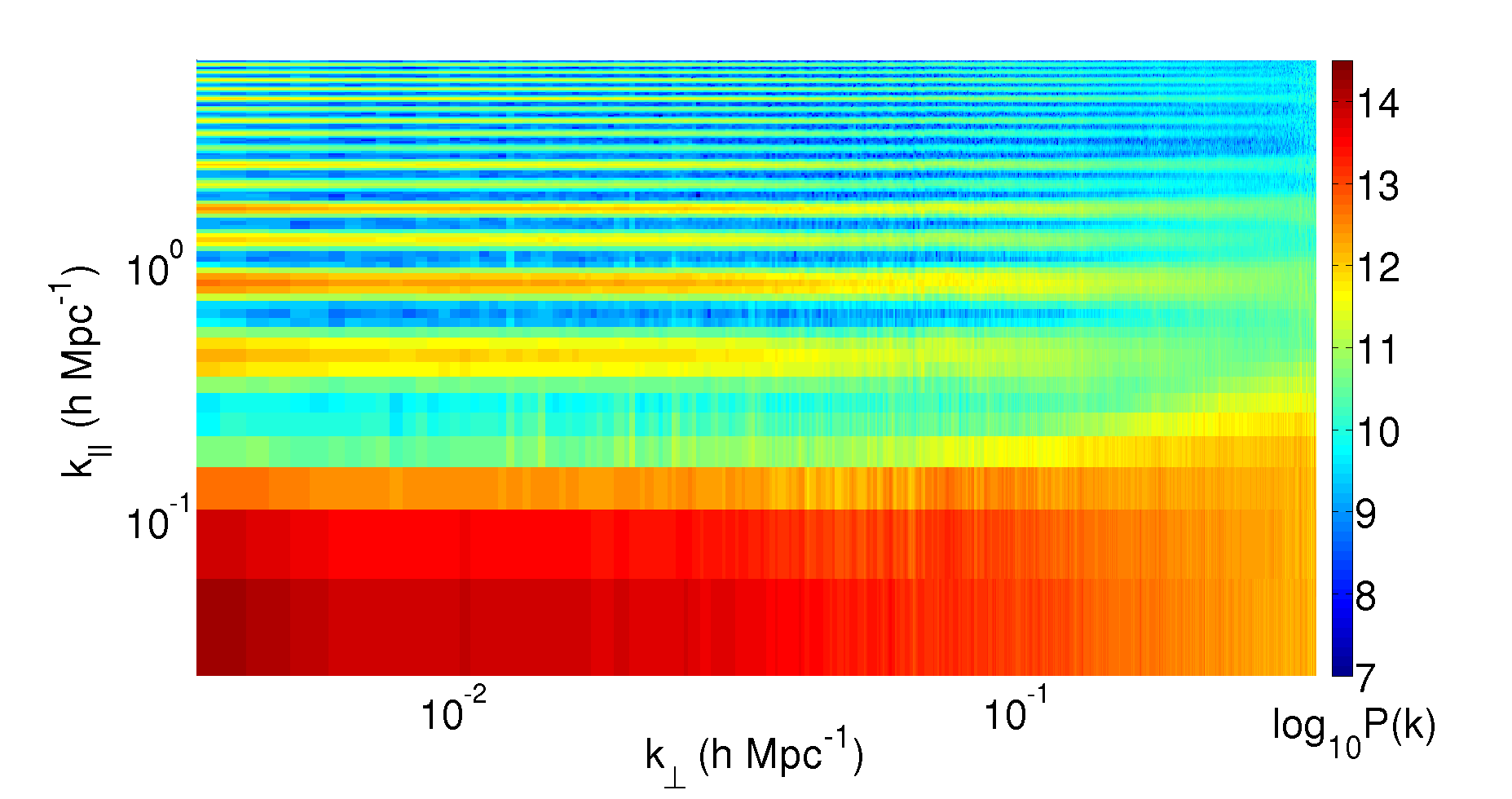}}\\
\subfloat[residual XX]{\includegraphics[width=0.5\textwidth]{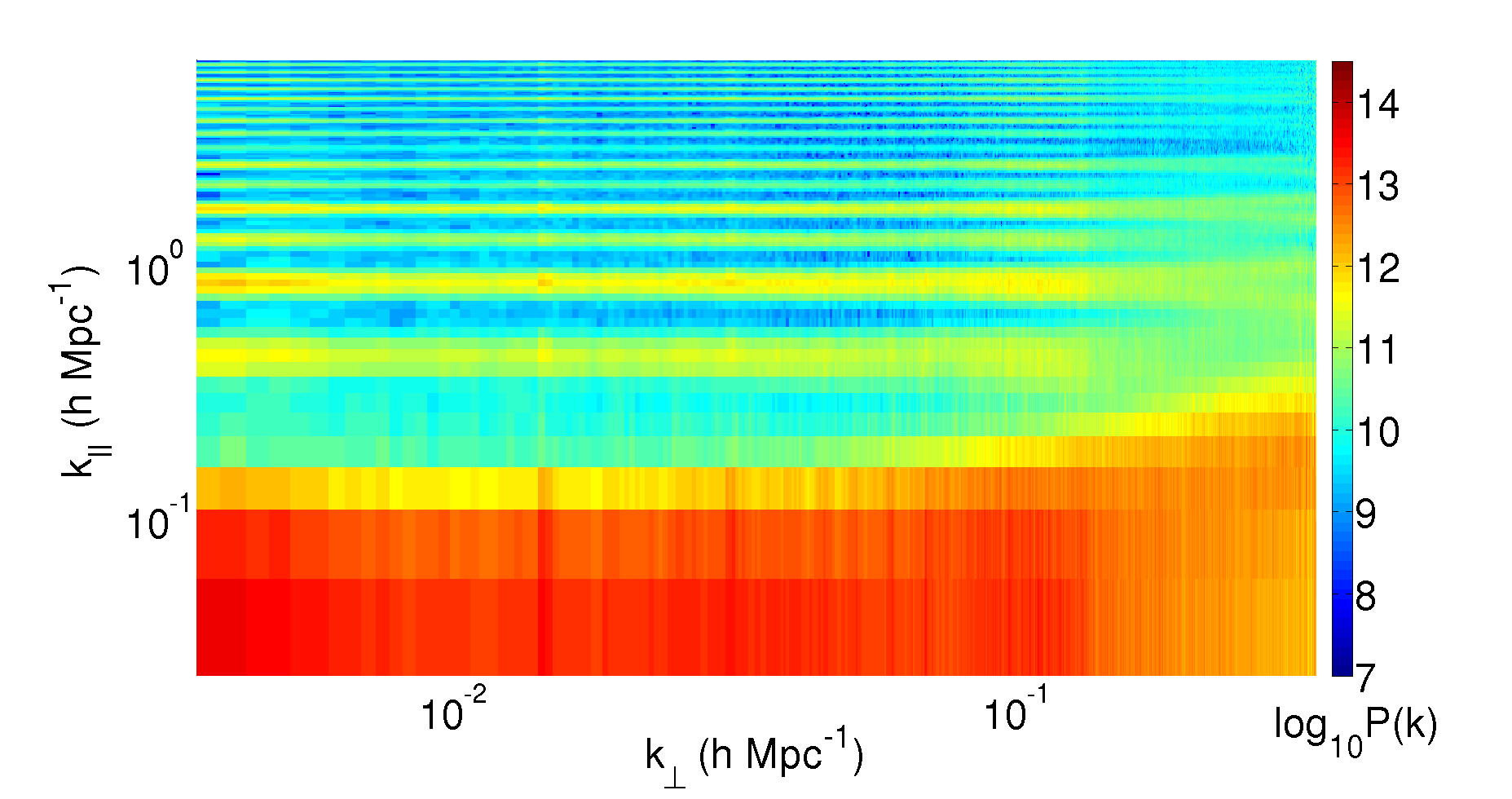}}
\subfloat[residual YY]{\includegraphics[width=0.5\textwidth]{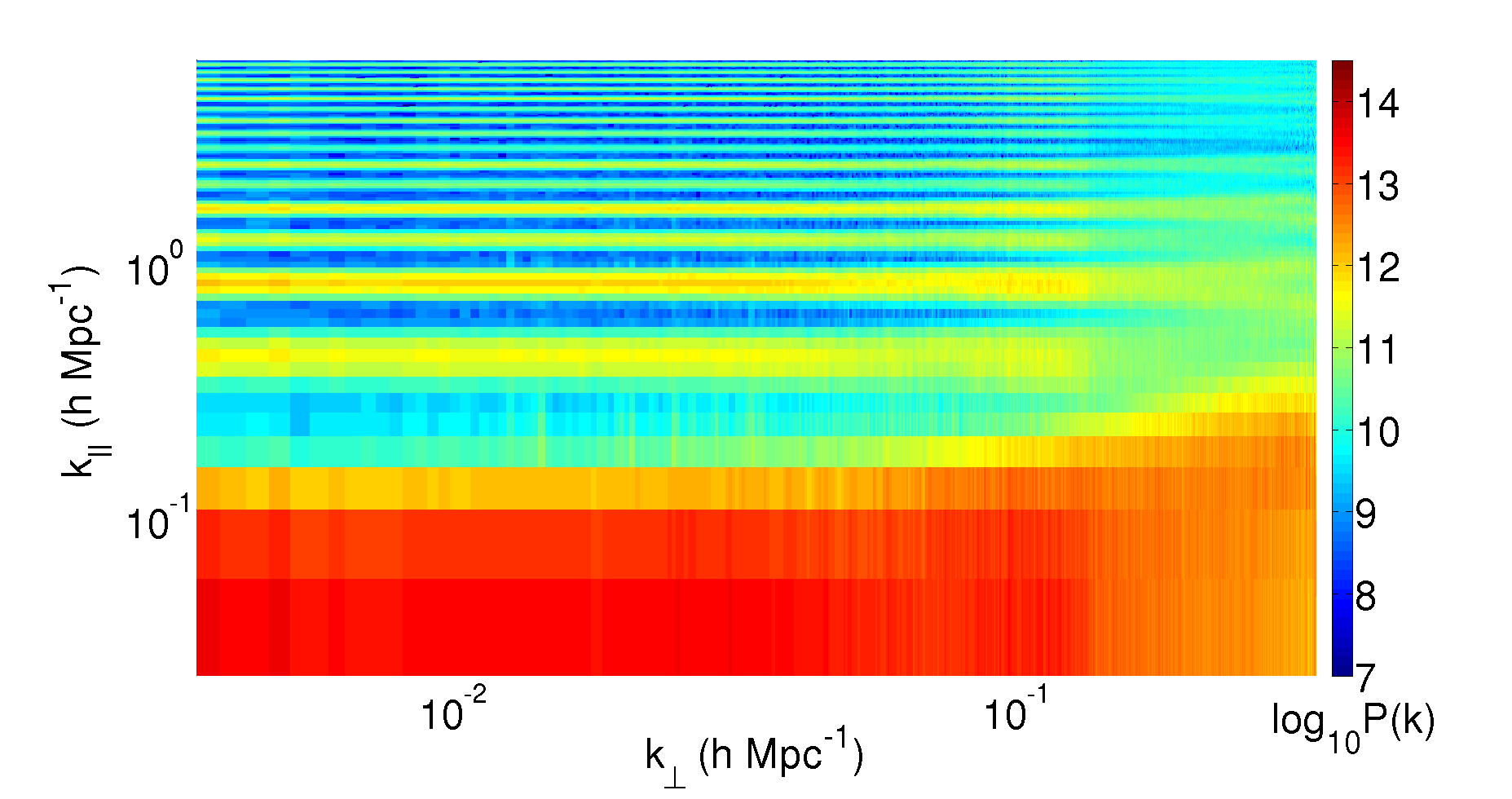}}\\
\caption{Two-dimensional power spectra for 3~hrs of data on the EoR1 field
 for XX \& YY polarization. The power is plotted as $log_{10}P$ where P is in units  $mK^2 (Mpc/h)^3$.}
\label{ps}
\end{figure*}

\begin{figure*}
\centering
\subfloat[XX]{\includegraphics[width=0.5\textwidth]{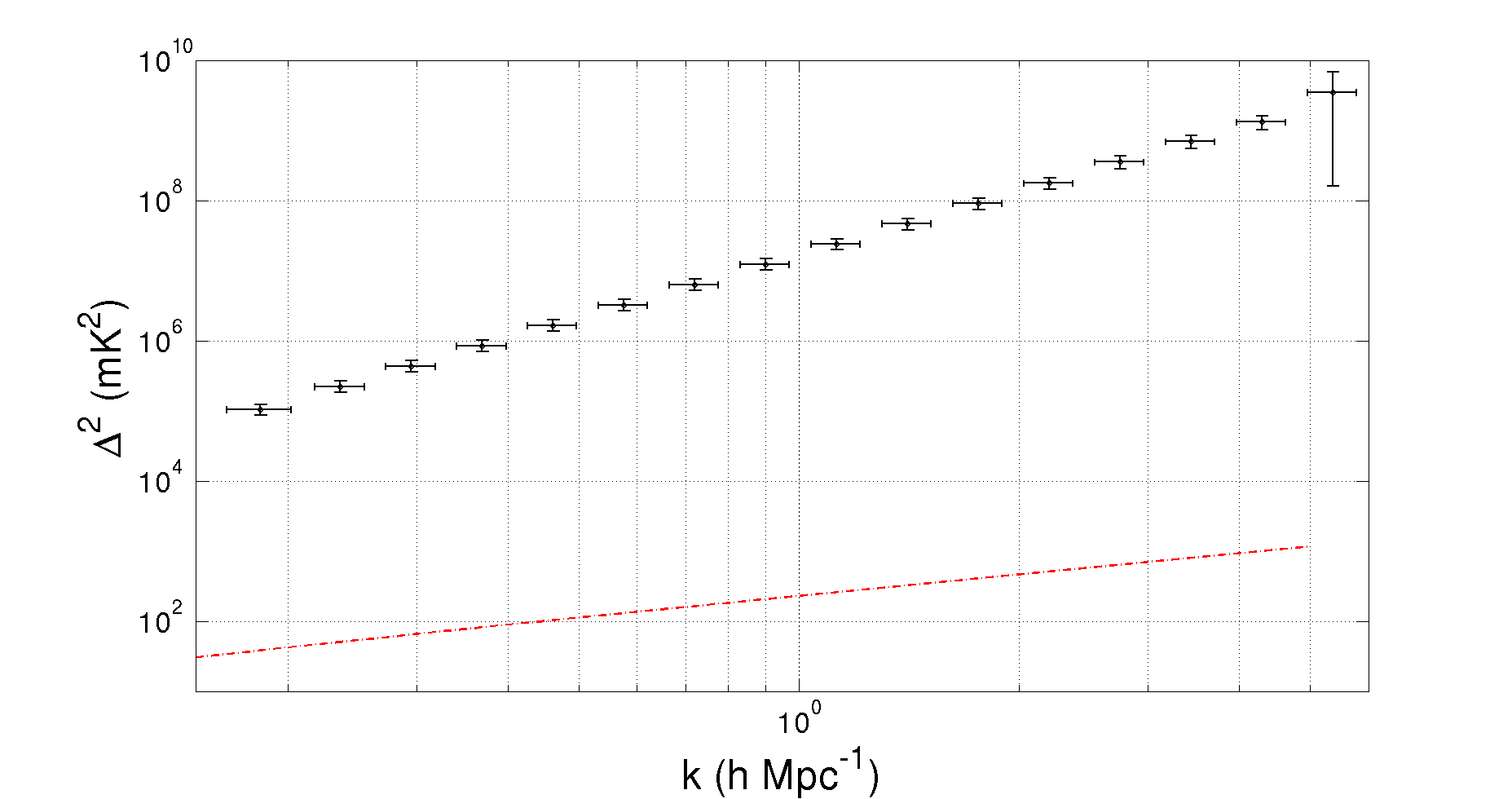}}
\subfloat[YY]{\includegraphics[width=0.5\textwidth]{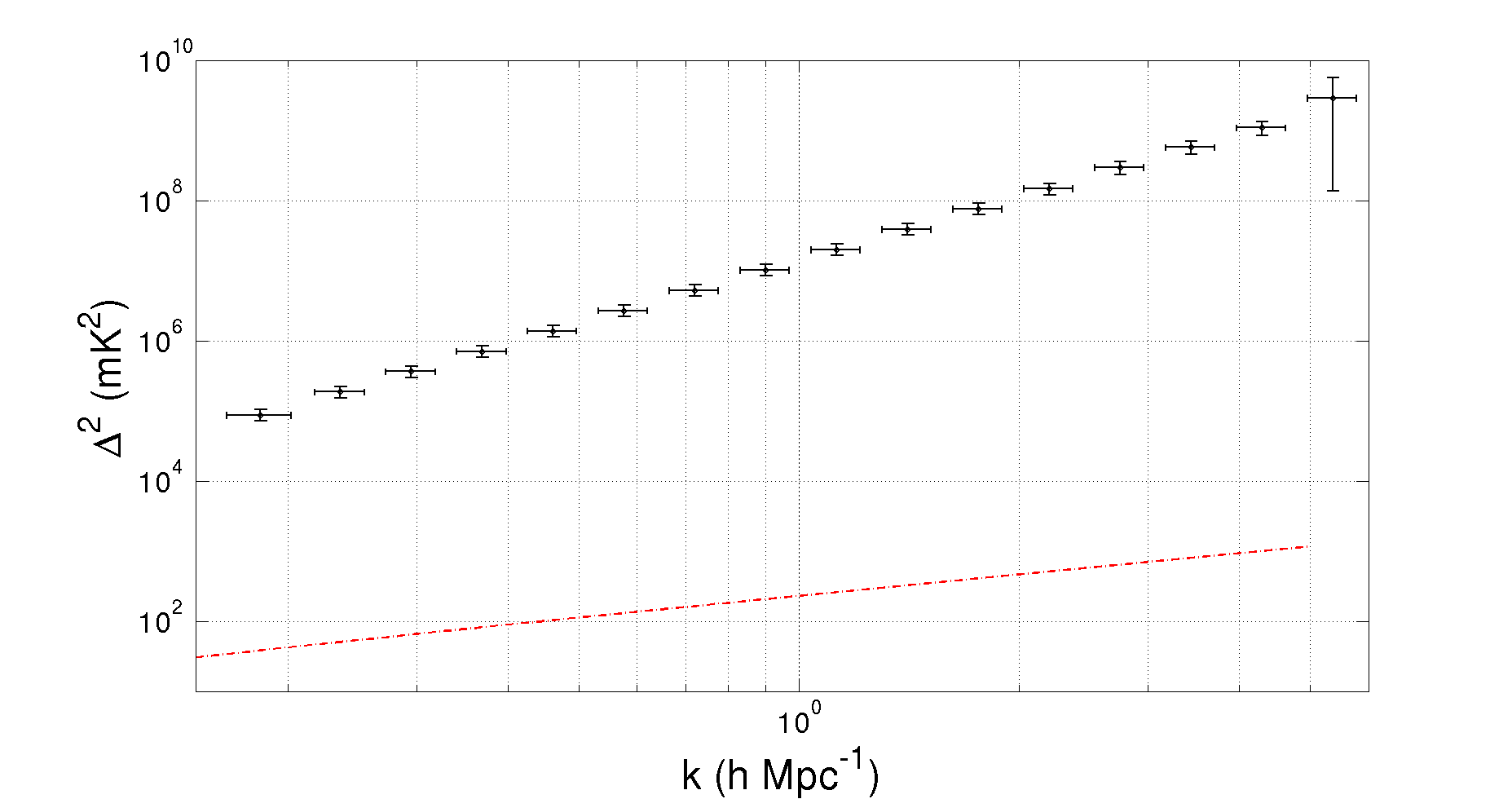}}\\
\caption{One dimensional power spectra  ($\Delta^2 = k^3 P(k)/(2\pi^2)$ in 
units $(mK)^2$) for the XX (Left panel)  and YY (Right panel) polarization are  shown  along with the errors
for each band. The dot-dashed (red) curve show the expected HI signal (\citet{Furlanetto et al}).}
\label{1d_power}
\end{figure*}

We first discuss discernible features in  the power spectra:
\begin{itemize}
\item[1.] In Fourier space the foreground contributions occupy a wedge shaped region (also called `foreground wedge') owing to the smooth spectral characteristics of foreground sources (\citet{wedge 1, wedge 2, Delay 1, Delay 2, Delay 3, Thyagrajan et al, Delay 4}). The region beyond the foreground wedge is expected to be free from the foreground contamination and dominated by thermal noise and expected HI signal. This foreground isolation approach is particularly useful for the MWA as it has low angular resolution. The strongest HI signals lie in the shortest baselines (low $k_\perp$ values) and it decreases rapidly with increasing $k_\perp$ values.  

 The upper  and lower  panels of  Figure~8 display  the dirty (calibrated with no foreground subtraction) and  the residual (clean components subtracted) power spectra, respectively. The Figure~\ref{ps} bear out the assumption that foregrounds
have smooth spectral characteristics as they are seen to form the `foreground wedge', this separation is in good agreement with the expectation  from foreground simulations (Figure~\ref{forenoise}). The first few $k_\parallel$ modes exhibit maximum foreground contributions, the $k_{\parallel}=0$ mode being the strongest. The amplitude at this mode is roughly $10^{14}\hbox{--}10^{15} \, \rm mk^2 \,  (Mpc/h)^3$ which is in good agreement  with the results of  other MWA EoR pipelines
 (\citet{Jacobs et al}). A clear decrement in power in the `foreground wedge' is visible in residual power spectra as compared to the dirty one.

\item[2.]  As described in the previous  section, MWA  has missing channels
on either side of   coarse bands of width 1.28~MHz. This leads to a periodicity of missing data across the frequency axis in visibility, the effect of which is reflected
 in the Fourier-transformed  power spectra as the horizontal bright lines at fixed $k_{\parallel}$.

\end{itemize}
 
Figure~\ref{1d_power} shows the 1-dimensional  power spectra; the 1-dimensional
power spectrum is obtained from regions that exclude   the foreground
wedge and bright coarse  bands  in  Figure~\ref{ps}. For computing the 1-d power spectrum, the foreground wedge and the bright coarse  horizontal bands of the 2-d power spectrum (Figure~\ref{ps})  are rejected.  More specifically, the channels corresponding to $k_\parallel \le  0.14 h Mpc^{-1}$ and $k_\perp > 0.1 h Mpc^{-1}$ are not considered. For each coarse band, the central  brightest channel along with two channels on either side are excluded.  The remaining contiguous regions are used in estimation of the 1-d power spectrum. For instance, for a given $k_\parallel$ that meets the criterion outlined above, all the cells that correspond to  $k_\perp < 0.1 h Mpc^{-1}$ are used for the computation of 1-d power spectrum.   The error
on the binned  power spectra are computed using a  scheme  outlined in Appendix~\ref{sec:powspecest1}.

 The 2-d power spectra obtained from data (Figure~\ref{ps}) can be compared with Figure~\ref{forenoise} that shows the expected power spectrum based  on
 simulations of foregrounds and noise. In particular, this comparison allows
us to assess the structure of coarse channels and the foreground wedge.
It also indicates the range of scales of the 2-d power spectrum.
 However,
Figure~\ref{forenoise} is based on a single realization of noise and a model of foregrounds
based on random distribution of point sources, and therefore a more detailed
comparison between the data and simulations is not possible. We shall return
to this detailed comparison in future work.

\section{Conclusions}

In this paper,  we propose a new method  to extract  the HI power spectrum from MWA visibility data in delay 
space. The proposed method  is applicable when a region is tracked using 
 imaging radio interferometers.

 One of the crucial factors in power spectrum estimation is how the $w$-term is dealt with within the pipeline. Our findings are that the $w$-term causes an effective shrinking of primary beam which reduces   the contribution of the 
 HI signal. We carefully model the HI signal by taking the $w$-term into account, the weights calculated are then applied to cross-correlate the measured visibilities. Moreover, the cross-correlation approach is particularly useful to minimize various systematics in the system.  We also model and account for the impact of 
changing intensity pattern in a tracking run. We find this effect to be 
sub-dominant to the $w$-term correction. 
 
We analyse three hours of MWA data from 
the EoR1 field, one of the field identified by the MWA community for EoR science. CASA has been used for calibration and to create foreground model using the clean components. Both the dirty (calibrated with no foreground subtraction) \& residual (foreground model subtracted) power spectrum in delay space are presented. Our  results (Figures~\ref{ps} and~\ref{1d_power}) are in good agreement with the    analyses  of other MWA EoR pipelines (\citet{Jacobs et al}).

In the future   we  plan  to apply the method proposed here for more 
integration time and, in particular, to  a longer  single tracking  run. 
The decorrelation caused by the $w$-term and the changing   intensity pattern would be more  dominant in the latter  case. This will allow us to 
 test the efficacy 
of our method for more extreme cases and might indicate the best possible 
way of detecting the HI signal from the epoch of reionization.

\section{Acknowlegements}
We thank the referee for useful comments which helped us to improve the paper.

This scientific work makes use of the Murchison Radio-astronomy Observatory, operated by CSIRO. We acknowledge the Wajarri Yamatji people as the traditional owners of the Observatory site. Support for the operation of the MWA is provided by the Australian Government Department of Industry and Science and Department of Education (National Collaborative Research Infrastructure Strategy: NCRIS), under a contract to Curtin University administered by Astronomy Australia Limited. We acknowledge the iVEC Petabyte Data Store and the Initiative in Innovative Computing and the CUDA Center for Excellence sponsored by NVIDIA at Harvard University.

\appendix

\section{Appendix A: Foregrounds and noise simulations}

\label{sec:foregrounds}

The primary contribution to foregrounds come from spectrally smooth 
 point and diffuse sources. They differ from the HI signal in both spatial
and spectral behaviour. However, it is the latter difference that allows us
to potentially isolate foregrounds from the HI signal in the power spectrum
estimation. 

To understand the impact of foregrounds in the data, we model them 
as a set of point sources. We note that  if  both the point and 
the diffuse sources have smooth spectra across the  instrumental bandwidth, their impact on the power spectra are similar and therefore point sources allow
us to capture adequately our ability to isolate foregrounds from the signal. 
In this section, for analytic work,  we assume $w =0$. We note without further proof that this assumption doesn't alter our main inferences. 


 For a set of point sources, the 
intensity distribution is given by:
\begin{equation}
I_\nu({\vec\theta}) = \sum_i F_{i\nu} \delta_{\scriptscriptstyle D}^2({\vec\theta} - {\vec\theta_i})
\end{equation}
Here $F_{i\nu}$ and ${\vec\theta_i}$  are the source fluxes and positions, respectively. This allows us to compute the  visibility  for a given baseline ${\bf u}_\nu$ and frequency $\nu$. 
\begin{equation}
V_\nu({\bf u}_\nu) = \sum_i F_{i\nu} \exp(i2\pi {\bf u}_\nu.{\vec\theta_i}) A_\nu({\vec\theta_i})
\label{vispoints}
\end{equation}
Here $A_\nu$ is the primary beam. As discussed earlier, we also 
define a visibility in the conjugate  space  by taking 
the Fourier transform with respect to $\nu$ (Eq.~(\ref{deftau})): $V_\tau({\bf u})$. Our aim
is to compute the correlation of this visibility:
\begin{equation}
\langle V_\tau ({\bf u}) V_{\tau'} ^* ({\bf u'}) \rangle = \sum_i \sum_j \int d\nu\int d\nu' F_{i\nu} F_{j\nu'} \exp\left(i2\pi\left[{\bf u}_\nu.\vec\theta_i - {\bf u}'_{\nu'}.\vec\theta_j +(\tau\nu-\tau'\nu') \right ] \right ) A_\nu(\vec\theta_i) A_{\nu'}(\vec\theta_j) 
\label{viscorpsou}
\end{equation}
Both the  source flux $F_\nu$ and primary beam $A_\nu(\vec\theta)$ are functions of frequency so it is difficult to analytically compute this expression. However, assuming 
smooth  and small variation of both of these quantities across the 
bandwidth, we can 
make meaningful analytic estimates; we verify this assumption 
from detailed simulations and the analysis of the data. The main frequency variation in this
case comes from the phase of the integral (the terms in the exponent)  and in particular from the 
change in the baseline length as frequency changes. We note here that multiple correlations  
are available to us for this analysis for different pairs of $\{\tau, \tau'\}$ and $\{{\bf u},{\bf u'}\}$. Here we assume $\tau = \tau'$. 

{\em Delay space---foreground wedge}: Here we  expand 
the same baseline in frequency space: ${\bf u'}_{\nu'} = {\bf u}_\nu + d{\bf u}_\nu/d\nu(\nu'-\nu)$.   In this case, ${\bf u}_\nu = \nu/\nu_0 {\bf u}_{\nu_0}$, where $\nu_0$ is some fixed frequency. Making the simplifying assumption that both point source fluxes and the primary beam are independent of frequency, Eq.~(\ref{viscorpsou})
can be analytically integrated. We further make coordinate transformation:
$x = (\nu'-\nu)/2$ and $y = (\nu'+\nu)/2$ and assuming $\nu'=\nu$ in 
 all the quantities except  those in the exponent containing their
 difference, which allows us to use $y \simeq \nu$: 
\begin{equation}
\langle V_\tau ({\bf u}) V_{\tau'} ^* ({\bf u'}) \rangle \simeq  \sum_i \sum_j \int dx\int dy F_{i\nu} F_{j\nu}\exp\left(i2\pi\left[{\bf u}_\nu.\vec\theta_i - {\bf u}_\nu.\vec\theta_j +2(d{\bf u}_\nu/d\nu.\vec\theta_i - \tau)x\right ] \right ) A_\nu(\vec\theta_i) A_\nu(\vec\theta_j) 
\label{viscorpsou1}
\end{equation}
Integrals over $x$ and $y$ are now separated which gives us:
\begin{equation}
\langle V_\tau ({\bf u}) V_{\tau'}^* ({\bf u'}) \rangle \propto \int dx \exp\left [-i 2\pi  (d{\bf u}_\nu/d\nu.\vec\theta_i - \tau)x \right ]
\end{equation}
As noted above, $d{\bf u}_\nu/d\nu = {\bf u}_{\nu_0}/\nu_0$, or it is independent of 
frequency. 
The integral in the equation is insignificant only when $\tau \simeq d{\bf u}/d\nu.\theta_i$. This linear relation between $\tau$ and the baseline ${\bf u}$ 
gives a region bounded by a  `wedge' in the $\tau$--$d{\bf u}/d\nu.\theta_i$ 
space for  a spatial  distribution of point sources  (e.g. see \citet{wedge 1, wedge 2, Delay 1, Delay 2, Delay 3, Thyagrajan et al, Delay 4}).

Another possible way to understand the nature of spectrally smooth 
foregrounds  is to first compute the correlation in the frequency
space. Using Eq.~(\ref{vispoints}), this gives us:
\begin{equation}
\langle V_\nu ({\bf u}_\nu) V_{\nu'} ^* ({\bf u'}_{\nu'}) \rangle = \sum_i \sum_j  F_{i\nu} F_{j\nu'}\exp\left(i2\pi\left[{\bf u}_\nu.\vec\theta_i - {\bf u}'_{\nu'}.\vec\theta_j \right ] \right ) A_\nu(\vec\theta_i) A_{\nu'}(\vec\theta_j) 
\label{viscorpsou2}
\end{equation}
Using   ${\bf u'}_{\nu'} = {\bf u}_\nu + \Delta\nu{\bf u}_{\nu_0}/\nu_0$  and substituting  into
Eq.~(\ref{viscorpsou2}),  and performing a single Fourier transform with respect to
$\Delta\nu$, we recover the main expected feature of the foreground 
`wedge' described above. Computationally, if the variation of other quantities with frequency, primary beam and  source fluxes, is neglected, this method is completely
equivalent to the one based on Eq.~(\ref{viscorpsou1}).

Even though we used a set of point sources, the main
inferences of the analysis also follow for diffuse sources. In particular, 
the frequency space integrals used to prove our case are exactly the same.

For our simulations, we assume  a set of point sources  isotropically 
distributed  with  fluxes above 1~Jy at 150 MHz. We construct this 
 flux distribution from radio source count at  1.4 GHz, which  is given by (\citet{Hopkins et al}):
\begin{equation}
\log \left [\frac {dn/ds}{S^{-2.5}} \right]= \sum_{i=0}^6 a_i \left [ \log \left(\frac {S}{mJy}\right) \right]^i 
\end{equation}
for flux range 0.05 mJy $\leq$ S $\leq$ 1000 mJy. The constants are $a_0=0.859, a_1=0.508, a_2=0.376, a_3=-0.049, a_4=-0.121, a_5=0.057, a_6=-0.008$.  We simulate sources over the entire hemisphere (nearly 15000~sources) to suitably
take into account the contribution from MWA primary beam  sidelobes. We extrapolate
the distribution to the frequencies of interest to us by assuming a spectral index $\alpha = -0.7$. 
\vspace{10 pt}

\subsection{Thermal noise}
Thermal noise is independent of the  baseline and depends on three parameters: system temperature, integration time and  the channel width.
The RMS of thermal noise associated with a visibility measurement  for channel width $\Delta\nu$ and integration time $\Delta t$ is: 
\begin{equation}
\sigma(\nu)= \frac{T_{\rm sys}}{K \sqrt{\Delta\nu \Delta t}}
\label{sigma_th}
\end{equation}
Here $T_{sys}$ and $K$ denote the system temperature and antenna gain respectively. For MWA, $K =A_{\rm eff}/(2k_B)$ with $A_{\rm eff} = 21.5 \, \rm m^2$ for MWA
at $\nu = 150 \, \rm MHz$ (\citet{MWA}). In our analysis we choose  $\Delta\nu=40$ kHz, $\Delta t=10$ seconds are very small compared to the frequency and time coherence of the signal (\citet{Drift paper}). The system temperature has two components: sky temperature (dominant source of noise at low frequency) and receiver temperature. We  consider $T_{\rm sys}=250 K$ for a single polarization  which is consistent with the reported system temperature  at 154.24~MHz for  the MWA pointing we consider in
this paper. 

It is fair to assume that the thermal noise for a radio interferometer follows a Gaussian statistics with zero mean. In our simulation,  we follow the same 
 pipeline used for analysing the  real data to estimate the thermal noise power.  We use the baseline distribution from the  observation with $u_{\rm max}=v_{\rm max}=250$. For every (u,v) point  the noise is drawn from a Gaussian distribution with zero mean and the  RMS  given by Eq.~(\ref{sigma_th}).

\begin{figure}
\centering
\includegraphics[width=0.5\textwidth]{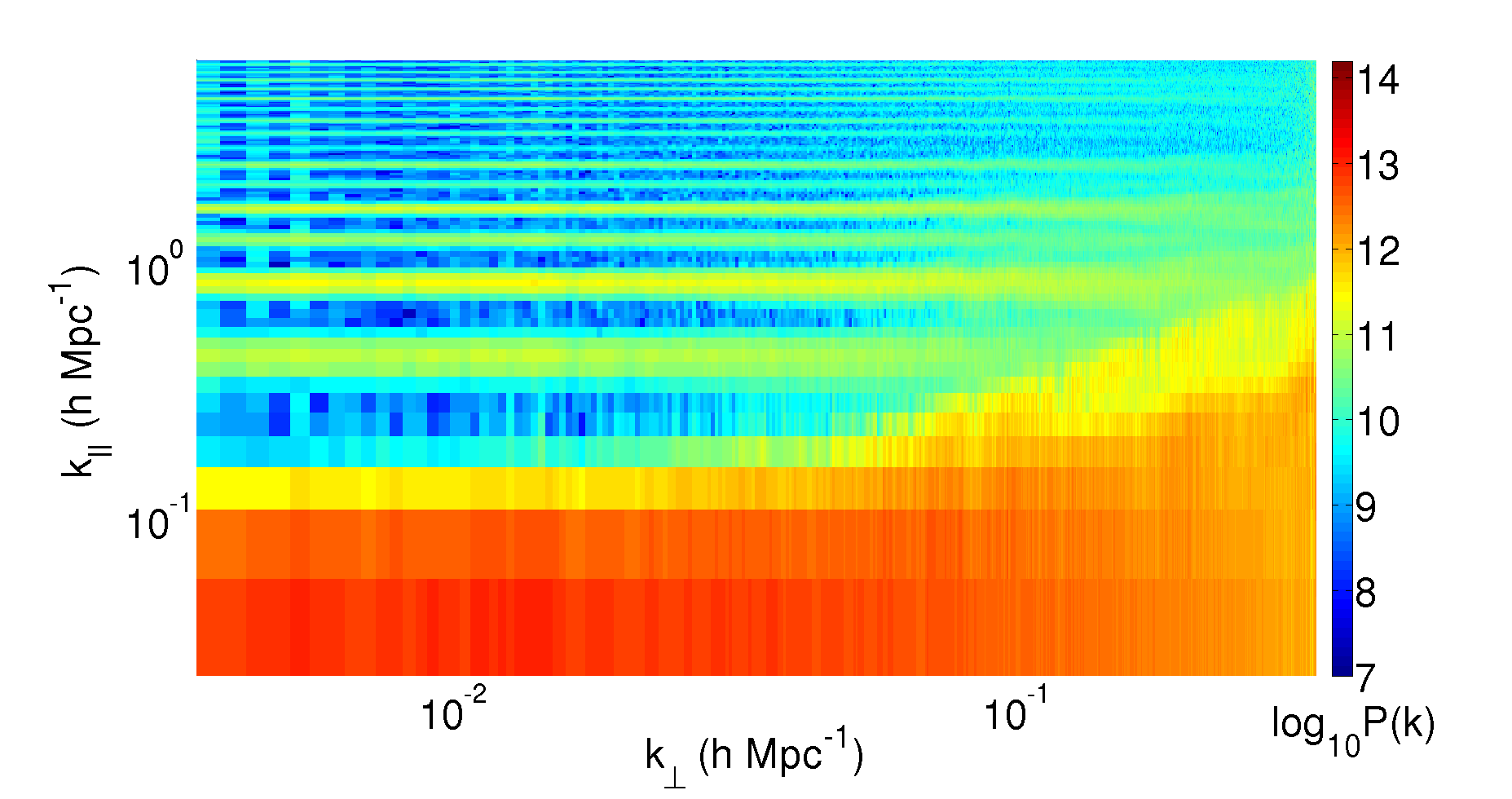}
\caption{The results of the simulation of  noise and foregrounds are 
shown. The power spectrum pipeline developed to analyse the data is also 
applied to  simulated foregrounds and noise for a three hour observational
run for a single polarization.}
\label{forenoise}
\end{figure}

\section{Appendix B: power spectrum estimation}

\label{sec:powspecest1}

As discussed in Section~\ref{sec:powspecest}, the power spectrum from the 
data is computed in two stages. First the power spectrum and its RMS is computed
for a single cell in which the HI signal is expected to be near coherent
and then an average is obtained across  cells assuming the HI signal
to be incoherent for different cells (Figure~\ref{schem}).

As noted in the text, the HI signal can be recovered
from a visibility cross-correlation by inverse weighing with ${\cal W}$.
We denote such a cross correlation: $\phat{S} \equiv V V/{\cal W}$; 
$\phat{S}$ is generally a complex number.   For optimal
averaging to get the lowest noise estimator,  one needs to sum over these cross-correlations 
by inverse weighing with the square of the RMS of each cross-correlation
$\sigma_i$. For pure noise,  $\sigma_i \propto 1/{\cal W}$ and  one can obtain Eq.~(\ref{powspecdef}). Notice that this estimator is invariant under an overall scaling of  $\sigma_i$. The error on power spectrum for each cell $\sigma_p$ is:
$\sigma_p^2 = \langle \phat{S}^2 \rangle - \langle \phat{S} \rangle^2$,
where the average is obtained optimally  from the data for all the cross-correlations. It can be shown that if each cross-correlation is assumed to be 
uncorrelated, as would be the case for pure noise, $\sigma_p \simeq (1/\sum\sigma_i^2)^{-1/2}$. Notice that if the RMS for all
the cross-correlations is the same, as would be the case if all measurements 
are equally weighted,    then this expression reduces to $\sigma_p = \sigma_i/\sqrt{N_c}$, where $N_c$ is the number of all the  cross-correlations within a  cell. 

This procedure yields an estimate of the power spectrum (Eq.~(\ref{powspecdef})) and its error $\sigma_p$ for each cell. 

For averaging over cells, we repeat the procedure described above by taking 
the estimated power spectrum for a cell as the signal and  
$\sigma_p$ as the weights. This allows us to estimate 2- and 1-dimensional 
power spectrum and its RMS.  For pure noise, the final error on the power spectrum 
is expected to approach: $\sigma_{\rm fin} = \sigma_p/\sqrt{N}$, where $N$ is the number of cells used for obtaining the average.

We briefly discuss some shortfalls of such a procedure. First, we do not 
construct the covariance matrix of the power spectrum estimator. We only 
estimate its diagonal terms, $\sigma_p^2$ and $\sigma_{\rm fin}^2$. This means 
that we are not able to assess the extent of cross-correlation between two neighbouring 
bins  in Figure~\ref{1d_power}. Such cross-correlation might contain important
information about systematic errors, foreground leakage, and  HI signal and 
noise cross correlation. 

Second, we do not include the HI signal in our estimation procedure. This is 
justified for the present work as the observed signal is clearly dominated by
noise and foreground residuals (Figure~\ref{1d_power}). We briefly assess
the impact of the HI signal for computing the error on the power spectrum. 

We assume the following  estimator for computing the power spectrum for a cell
and consider the contribution of only the HI signal:
\begin{equation}
 \phat{S} = {1\over N_c}  \sum_{ij} {V_i V_j \over {\cal W}_{ij}}
\end{equation}
As noted above, this estimator allows us to recover the HI signal.  The 
subscripts $ij$ correspond to a pair for visibilities and the sum is 
carried over all the cross-correlations. After further computation, we obtain
the error on the signal:
\begin{equation}
\Delta\phat{S}^2 = {\phat{S}^2 \over N_c^2}\sum_{kl}\sum_{ij} \left[{{\cal W}_{ik} {\cal W}_{jl} + {\cal W}_{il} {\cal W}_{jk} \over {\cal W}_{ij} {\cal W}_{kl}} \right ] 
\end{equation}
If all the weights are unity this reduces to the usual cosmic variance 
expression: $\Delta\phat{S} = \sqrt{2} \phat{S}$. Even though this term
is negligible for our purposes, this would need to be included for longer
integration times.

\end{document}